\documentstyle[aps,prd,psfig]{revtex}
\newcommand{\bm}[1]{{\mbox{\boldmath$#1$}}}

\def\Quadrat#1#2{{\vcenter{\hrule height #2
  \hbox{\vrule width #2 height #1 \kern#1
    \vrule width #2}
  \hrule height #2}}}
\def\dAl{\mathop{\kern 1pt\hbox{$\Quadrat{8pt}{0.4pt}$} \kern1pt}}
\begin{document}

\author{Sergei Kopeikin and Bahram Mashhoon}
\address{Department of Physics \& Astronomy,
University of Missouri-Columbia,\\ Columbia, Missouri 65211, USA}
\title{Gravitomagnetic Effects in the Propagation of Electromagnetic Waves
in Variable Gravitational Fields of
Arbitrary-Moving and Spinning Bodies}
\maketitle
\tableofcontents
\date{\today}
\baselineskip=20pt
\newpage
\begin{abstract}
Propagation of light in the gravitational field of self-gravitating spinning bodies
moving with arbitrary velocities is discussed. The gravitational field is assumed to be ``weak" everywhere. Equations of motion of a light ray are solved in the first post-Minkowskian approximation that is linear with respect to the universal gravitational constant $G$. We do not restrict ourselves with the approximation of gravitational lens so that the solution of light geodesics is applicable for arbitrary locations of source of light and observer. This formalism is applied for studying corrections to the Shapiro time delay in binary pulsars caused by the rotation of pulsar and its companion. We also derive the correction to the light deflection angle caused by rotation of gravitating bodies in the solar system (Sun, planets) or a gravitational lens. The gravitational shift of frequency due to the combined translational and rotational motions of light-ray-deflecting bodies is analyzed as well. We give a general derivation of the formula describing the relativistic rotation of the plane of polarization of electromagnetic waves (Skrotskii effect). This formula is valid for arbitrary translational and rotational motion of gravitating bodies and greatly extends the results of previous researchers. Finally, we discuss the Skrotskii effect for gravitational waves emitted by localized sources such as a binary system. The theoretical results of this paper can be applied for studying various relativistic effects in microarcsecond space astrometry and developing corresponding algorithms for data processing in space astrometric missions such as FAME, SIM, and GAIA.
\end{abstract}
\section{Introduction}

The influence of gravitation on the propagation of electromagnetic rays has been treated by many authors since Einstein first calculated the relativistic deflection of light by a spherically symmetric mass \cite{ein}. Currently there is much interest in several space missions dedicated to measuring astrometric positions, parallaxes, and proper motions of stars and quasars with an accuracy approaching 1 microarcsecond ($\mu$arcsec) \cite{mission}. Thus progress in observational techniques has made it necessary to take into account the fact that electromagnetic rays are deflected not only by the monopole gravitoelectric field of light-deflecting bodies but also their spin-gravitomagnetic and quadrupolar gravitoelectric fields. Propagation of light in the stationary gravitational fields of rotating oblate bodies with their centers of mass at rest is well known (see, e.g., \cite{chandra} --\cite{kop1}, and references therein). Quite recently significant progress has been achieved in solving the problem of propagation of light rays in the field of an isolated source (like a binary star) that has a time-dependent quadrupole moment and emits quadrupolar gravitational waves \cite{KSGE}. However, the translational motion of the source of gravitational waves was not taken into account in \cite{KSGE}; in effect, the source was assumed to be at rest.

We would like to emphasize the point that most gravitational sources are not in general static and move with respect to the observer in a variety of ways. As revealed by approximate estimates \cite{BKK}, the precision of planned astrometric space missions \cite{mission}, binary pulsar timing tests of general relativity \cite{T}, very long baseline interferometry \cite{Ma} --\cite{Fey}, etc., necessitates the development of more general methods of integration of equations of light propagation that could account for such translational motions. Indeed, non-stationary sources emit gravitational waves that weakly perturb the propagation of electromagnetic signals. The cumulative effect of such gravitational waves may be quite large and, in principle, could be detectable \cite{grwave}. It is necessary to know how large this influence is and whether one can neglect it or not in current and/or planned astrometric observations and experimental tests of general relativity. When considering propagation of light in the field of moving bodies it is also worth keeping in mind that gravitational interaction propagates with the speed of light in linearized general relativity \cite{will} --\cite{w}. This retarded interaction has important consequences and can play a crucial role in the theoretical prediction of secondary relativistic effects in time delay, light deflection, polarization of light, etc.

The first crucial step in solving the problem of propagation of electromagnetic rays in the retarded gravitational field of arbitrary-moving bodies has been taken recently in  \cite{KS}. The new formalism allows one to obtain a detailed description of the light ray trajectory for unrestricted locations of the source of light and observer and to make unambiguous predictions for possible relativistic effects. An important feature of the formalism is that it is based on the post-Minkowskian solution \cite{bertpleb} --\cite{dmur} of the linearized Einstein field equations. Thus the amplitude of gravitational potentials is assumed to be small compared to unity, but there are no a priori restrictions on velocities, accelerations, etc., of the light-deflecting bodies. In this way the retarded character of the gravitational field is taken into account in the linear approximation. This is in contrast to the post-Newtonian approximation scheme \cite{fock} --\cite{bds}, which assumes that velocities of light-deflecting bodies must be small with respect to the speed of light. Such a treatment destroys the causal character of a gravitational null cone and makes the gravitational interaction appear to propagate instantaneously at each step of the iteration procedure. As proved in \cite{KS}, this is the reason why the post-Newtonian metric gives correct answers for the time delay and light deflection angle only for a very restricted number of physical situations. As a rule, more subtle (secondary) effects in the propagation of light rays in time-dependent gravitational fields are not genuinely covered in the post-Newtonian approach, whereas the post-Minkowskian light-propagation formalism gives unique and unambiguous answers.

In the present paper, we extend the light-propagation formalism
developed in \cite{KS} in order to be able to include the
relativistic effects related to the gravitomagnetic field produced
by the translational velocity-dependent terms in the metric tensor
as well as spin-dependent terms due to light-deflecting bodies. We
shall start from the consideration of the energy-momentum tensor
of moving and spinning bodies (section 2). Then, we solve the
Einstein field equations in terms of the retarded {\it
Li\'enard-Wiechert} potentials (section 3), describe the light-ray
trajectory in the field of these potentials (section 4), and then
calculate the deflection angle and time delay in the propagation
of electromagnetic rays through the system of arbitrary-moving and
spinning particles (section 5). We pay particular attention to the
calculation of gravitomagnetic effects in ray propagation in
pulsar timing and astrometry (section 6). Moreover, we discuss the
relativitic effect of the rotation of the plane of polarization of
electromagnetic rays (Skrotskii effect \cite{wecal} --\cite{kl11})
in the gravitomagnetic field of the above-mentioned system of
massive bodies (section 7). This effect may be important for a
proper interpretation of a number of astrophysical phenomena that
take place in the accretion process of X-ray binaries
\cite{star} --\cite{chen} and/or supermassive black holes that may
exist in active galactic nuclei. Finally, we derive an exact
expression for the rotation of the plane of polarization of light
caused by the quadrupolar gravitational waves emitted by localized
sources (section 8). The treatment of the Skrotskii effect given
in the final section may be important for understanding the
effects of cosmological gravitational waves on the anisotropy of
cosmic microwave background (CMB) radiation at different scales
\cite{kra} --\cite{turn}. Details related to the calculation of
integrals along the light-ray trajectory are relegated to the
appendices.

\section{Energy-Momentum Tensor of Spinning Body}

We consider an ensemble of $N$ self-gravitating bodies possessing
mass and spin; higher-order multipole moments are neglected for
the sake of simplicity; for their influence on light propagation
via the mathematical techniques of the present paper see
\cite{quad}. Propagation of light in the gravitational field of
arbitrary-moving point-like masses has been studied in \cite{KS}.
The energy-momentum tensor $T^{\alpha\beta}$ of a spinning body is
given
\begin{eqnarray}
\label{1}
T^{\alpha\beta}(t,{\bf x})&=&T^{\alpha\beta}_M(t,{\bf x})+
T^{\alpha\beta}_S(t,{\bf x})\;,
\end{eqnarray}
where $T^{\alpha\beta}_M$ and $T^{\alpha\beta}_S$ are pieces of
the tensor generated by respectively the mass and spin of the
body, and $t$ and ${\bf x}$ are coordinate time and spatial
coordinates of the underlying inertial coordinate system. In the
case of several spinning bodies the total tensor of
energy-momentum is a linear sum of tensors of the form (\ref{1})
corresponding to each body. Therefore, in the linear approximation
under consideration in this paper, the net gravitational field is
simply a linear superposition of the field due to individual
bodies.

In equation (\ref{1}), $T^{\alpha\beta}_M$ and $T^{\alpha\beta}_S$ are defined in
terms of the Dirac $\delta$-function \cite{1} --\cite{3} as follows
\cite{4} --\cite{6}
\begin{eqnarray}
\label{2} T^{\alpha\beta}_M(t,{\bf
x})&=&\displaystyle{\int_{-\infty}^{+\infty}} p^{(\alpha}u^{\beta
)}(-g)^{-1/2} \delta(t-z^0(\eta))\;\delta({\bf x}-{\bf
z}(\eta))d\eta\;,
\end{eqnarray}
\begin{eqnarray}
\label{3} T^{\alpha\beta}_S(t,{\bf
x})&=&-{\bigtriangledown}_{\gamma}\displaystyle{\int_{-\infty}^{+\infty}}
S^{\gamma(\alpha}u^{\beta )}(-g)^{-1/2}
\delta(t-z^0(\eta))\;\delta({\bf x}-{\bf z}(\eta))d\eta\;,
\end{eqnarray}
where $\eta$ is the proper time along the world-line of the body's
center of mass, ${\bf z}(\eta)$ are spatial coordinates of the
body's center of mass at proper time $\eta$,
$u^\alpha(\eta)=u^0(1,v^i)$ is the 4-velocity of the body,
$u^0(\eta)=(1-v^2)^{-1/2}$, $(v^i)\equiv{\bf v}(\eta)$ is the
3-velocity of the body in space, $p^\alpha(\eta)$ is the body's
linear momentum (in the approximation neglecting rotation of
bodies $p^\alpha(\eta)=mu^\alpha$, where $m$ is the invariant mass
of the body), $S^{\alpha\beta}(\eta)$ is an antisymmetric tensor
representing the body's spin angular momentum attached to the
body's center of mass, ${\bigtriangledown}_{\gamma}$ denotes
covariant differentiation with respect to metric tensor
$g_{\alpha\beta}$, and $g={\rm det}(g_{\alpha\beta})$ is the
determinant of the metric tensor.

The definition of $S^{\alpha\beta}$ is arbitrary up to the choice
of a spin supplementary condition that is chosen as follows
\begin{eqnarray}
\label{cond}
S^{\alpha\beta}u_{\beta}=0\;;
\end{eqnarray}
this constraint is consistent with neglecting the internal
structure of the bodies involved, so that we deal in effect with
point-like spinning particles \cite{m71} --\cite{m75}. We introduce
a spin vector $S^\alpha$ which is related to the spin tensor by
the relation
\begin{eqnarray}
\label{spvec}
S^{\alpha\beta}&=&\eta^{\alpha\beta\gamma\delta}u_\gamma S_\delta\;,
\end{eqnarray}
where $\eta^{\alpha\beta\gamma\delta}$ is the Levi-Civita tensor
related to the completely antisymmetric Minkowskian tensor
$\epsilon_{\alpha\beta\gamma\delta}$ \cite{7} as follows
\begin{eqnarray}
\label{levi}
\eta^{\alpha\beta\gamma\delta}=-(-g)^{-1/2}\epsilon_{\alpha\beta\gamma\delta}\;,
\quad\quad\quad\quad\quad
\eta_{\alpha\beta\gamma\delta}=(-g)^{1/2}\epsilon_{\alpha\beta\gamma\delta}\;,
\end{eqnarray}
where $\epsilon_{0123}=+1$.
The spin vector is orthogonal to $u^\alpha$ by definition so that the identity
\begin{eqnarray}
\label{qera}
S^\alpha u_\alpha&\equiv&0\;
\end{eqnarray}
is always valid; this fixes the one remaining degree of freedom $S^0$. Hence, the 4-vector
$S^\alpha$ has only three independent spatial components.

The dynamical law ${\bigtriangledown}_{\beta} T^{\alpha\beta}=0$,
applied to equations (\ref{1}) -(\ref{3}), leads to the equations
of motion of a spinning particle in a gravitational field. Indeed,
using a theorem of L. Schwartz that a distribution with a simple
point as support is a linear combination of Dirac's delta function
and a finite number of its derivatives, one can develop the theory
of the motion of point-like test bodies with multipole moments in
general relativity \cite{taub}. In this way, it can be shown in
particular that for a ``pole-dipole" particle under consideration
in this paper, one is led uniquely \cite{taub} to the
Mathisson-Papapetrou equations with the Pirani supplementary
condition ({\ref{cond}).

In what follows, we focus only on effects that are produced by the
spin and are linear with respect to the spin and Newton's
gravitational constant (the first post-Minkowskian approximation)
in the underlying asymptotically inertial global coordinate
system. The effects produced by the usual point-particle piece of
the energy-momentum tensor have already been studied in \cite{KS}.
Let us in what follows denote the components of the spin vector in
the frame comoving with the body as ${\cal J}^\alpha=(0,{\bm{\cal
J}})$. In this frame the temporal component of the spin vector
vanishes as a consequence of (\ref{qera}) and after making a
Lorentz transformation from the comoving frame to the underlying
``inertial" frame, we have in the post-Minkowskian approximation
\begin{equation}
\label{tgh}
S^0=\gamma({\bf v}\cdot{\bm{\cal J}})\;,\quad\quad\quad\quad
S^i={\cal J}^i+\frac{\gamma-1}{v^2}({\bf v}\cdot{\bm{\cal J}})v^i\;,
\end{equation}
where $\gamma\equiv(1-v^2)^{-1/2}$ and ${\bf v}=(v^i)$ is the
velocity of the body with respect to the frame at rest.

\section{Gravitational Field Equations and Metric Tensor}

The metric tensor in the linear approximation can be written as
\begin{eqnarray}
\label{2a}
g_{\alpha\beta}(t, {\bf x})&=&\eta_{\alpha\beta}+h_{\alpha\beta}(t, {\bf
x})\;,
\end{eqnarray}
where $\eta_{\alpha\beta}={\rm diag}(-1,+1,+1,+1)$ is the Minkowski metric of
flat space-time and the metric perturbation
$h_{\alpha\beta}(t, {\bf x})$ is a function of time and
spatial coordinates \cite{3}. We split the metric perturbation into two pieces
$h^{\alpha\beta}_M$ and $h^{\alpha\beta}_S$
that are linearly independent in the first post-Minkowskian approximation;
that is,
\begin{eqnarray}
\label{chop}
h^{\alpha\beta}&=&h^{\alpha\beta}_M+h^{\alpha\beta}_S\;.
\end{eqnarray}
Thus, the solution for the
each piece can be found from the Einstein field equations with the corresponding
energy-momentum tensor.

The point-particle piece $h^{\alpha\beta}_M$ of the metric tensor
has already been discussed in \cite{KS} and is given by
\begin{eqnarray}
\label{2aa}
h^{\alpha\beta}_M&=&4m\sqrt{1-v(s)^2}\;\frac{u(s)^\alpha u(s)^\beta+{1\over2}
\eta^{\alpha\beta}}{r(s)-{\bf v}(s)\cdot{\bf r}(s)}\;,
\end{eqnarray}
where $r(s)=|{\bf r}(s)|$, ${\bf r}(s)={\bf x}-{\bf z}(s)$, and both coordinates
${\bf z}$ and
velocity ${\bf v}$ of the body are assumed to be time dependent
and calculated at the retarded moment of time
$s$ defined
by the light-cone equation
\begin{eqnarray}
\label{fgh2}
s+|{\bf x}-{\bf z}(s)|=t\;,
\end{eqnarray}
that has a vertex at the space-time point $(t,{\bf x})$ and
describes the propagation of gravitational field \cite{propa} on
the unperturbed Minkowski space-time. The solution of this equation
gives the retarded time $s$ as a function of coordinate time $t$
and spatial coordinates ${\bf x}$, that is $s=s(t,{\bf x})$.

As for
$h^{\alpha\beta}_S$, it can be found by solving the field equations
that are given
in the first post-Minkowskian approximation and in the harmonic
gauge \cite{9} as follows
\begin{eqnarray}
\label{3a}
\dAl h_S^{\alpha\beta}(t, {\bf x})&=&-16\pi \left[T_S^{\alpha\beta}(t, {\bf x})-
\frac{1}{2}\eta^{\alpha\beta}\;T_{S\;\lambda}^{\lambda}(t, {\bf
x})\right]\;.
\end{eqnarray}
Taking into account the orthogonality condition (\ref{cond}), the field equations
(\ref{3a}) assume the form
\begin{eqnarray}
\label{fgh} \dAl h_S^{\alpha\beta}(t, {\bf
x})&=&16\pi{\partial}_{\gamma}\int^{+\infty}_{-\infty}d\eta
\left[S^{\gamma(\alpha}(\eta)u^{\beta)}(\eta)
\delta(t-z^0(\eta))\;\delta({\bf x}-{\bf z}(\eta))\right]\;,
\end{eqnarray}
where we have replaced the covariant derivative ${\bigtriangledown}_{\gamma}$
with a simple partial derivative ${\partial}_{\gamma}=
\partial/\partial x^\gamma$ and $u^\alpha=\gamma
(1, v^i)$, where $\gamma=(1-v^2)^{-1/2}$ is the Lorentz factor.
The solution of these equations is given by the {\it Li\'enard-Wiechert} spin-dependent
potentials
\begin{eqnarray}
\label{fgh1}
h_S^{\alpha\beta}(t, {\bf x})&=&-4{\partial}_{\gamma}
\left[\frac{S^{\gamma(\alpha}(s)u^{\beta)}(s)}{r(s)-{\bf v}(s)\cdot{\bf r}(s)}
\right]\;.
\end{eqnarray}

If we restricted ourselves with the linear approximation of
general relativity, the sources under consideration here would
have to be treated as a collection of free non-interacting
spinning test particles each moving with arbitrary constant speed
with its spin axis pointing in an arbitrary fixed direction. This
simply follows from the Mathisson-Papapetrou equations in the
underlying inertial coordinate system. Thus, in such a case
carrying out the differentiation in (\ref{fgh1}), we arrive at
\begin{eqnarray}
\label{aqw}
h_S^{\alpha\beta}(t, {\bf x})&=&4(1-v^2)\;\frac{r_\gamma S^{\gamma(\alpha}u^{\beta)}}{[r(s)-{\bf v}\cdot{\bf r}(s)]^3}\;,
\end{eqnarray}
where ${\bf v}$ and $S^{\alpha\beta}$ are treated as constants and
we define $r^\alpha=(r, {\bf r})$. However, we may consider
(\ref{2a}) as expressing the first two terms in a post-Minkowskian
perturbation series; that is, at some ``initial" time we turn on
the gravitational interaction between the particles and keep track
of terms in powers of the gravitational constant only, without
making a Taylor expansion with respect to the ratio of the
magnitudes of the characteristic velocities of bodies to the speed
of light. The development of such a perturbative scheme involves
many specific difficulties as discussed in
\cite{bertpleb} --\cite{dmur}. In this approach one may relax the
restrictions on the body's velocity and spin and think of ${\bf
v}$ and $S^{\alpha\beta}$ in (\ref{aqw}) as arbitrary functions of
time. In this paper as well as \cite{KS}, we limit our
considerations to the first-order equations (\ref{chop}),
(\ref{2aa}), and (\ref{aqw}); however, in the {\it application} of
these results to the problem of ray propagation, we let ${\bf v}$
and $S^{\alpha\beta}$ in the {\it final} results based on
(\ref{2aa}) and (\ref{aqw}) be time dependent as required in the
specific astrophysical situation under consideration. Therefore,
the consistency of the final physical results with the
requirements of our approximation scheme must be checked in every
instance.

The metric tensor given by equations (\ref{chop}), (\ref{2aa}),
and (\ref{aqw}) can be used to solve the problem of propagation of
light rays in the gravitational field of arbitrary-moving and
spinning bodies.

\section{Propagation Laws for Electromagnetic Radiation}

The general formalism describing the behavior of electromagnetic
radiation in an arbitrary gravitational field is well known \cite{genfor}.
A high-frequency electromagnetic wave is defined as an approximate
solution of the Maxwell equations of the form
\begin{equation}
\label{ff1}
F_{\alpha\beta}={\rm Re}\{A_{\alpha\beta}\exp(i\varphi)\}\;,
\end{equation}
where $A_{\alpha\beta}$ is a slowly varying function of position
and $\varphi$ is a rapidly varying phase of the electromagnetic
wave. From equation (\ref{ff1}) and Maxwell's equations one
derives the following results (for more details see \cite{MTW} --
\cite{mash87} and Appendix A). The electromagnetic wave vector
${l}_\alpha=\partial\varphi/\partial x^\alpha$ is real and null,
that is $g^{\alpha\beta}{l}_\alpha {l}_\beta=0$. The curves
$x^\alpha=x^\alpha(\lambda)$ that have
${l}^\alpha=dx^\alpha/d\lambda$ as a tangent vector are null
geodesics orthogonal to the surfaces of constant electromagnetic
phase $\varphi$. The null vector ${l}^\alpha$ is parallel
transported along itself according to the null geodesic equation
\begin{equation}
\label{p0}
{l}^\beta\nabla_\beta{l}^\alpha=0\;.
\end{equation}
The equation of the parallel transport (\ref{p0}) can be expressed
as
\begin{equation}
\label{ff2}
{d{l}^\alpha\over d\lambda}+\Gamma^\alpha_{\beta\gamma}{l}^\beta
{l}^\gamma=0\;,
\end{equation}
where $\lambda$ is an affine parameter. The
electromagnetic field tensor $F_{\alpha\beta}$ is a null field satisfying
$F_{\alpha\beta}{l}^\beta=0$ whose propagation law in an
arbitrary empty space-time is
\begin{equation}
\label{p1}
D_\lambda F_{\alpha\beta}+\theta F_{\alpha\beta}=0\;,\quad\quad\quad\quad\quad
D_\lambda\equiv {D\over D\lambda}={dx^\alpha\over d\lambda} \nabla_\alpha\;,
\end{equation}
where $\theta={1\over 2}\nabla_\alpha{l}^\alpha$ is the
expansion of the null congruence ${l}_\alpha$.

Let us now construct a null tetrad $({l}^\alpha, n^\alpha,
m^\alpha, \overline{m}^\alpha)$, where the bar indicates complex
conjugation, ${n}_\alpha l^\alpha=-1$ and $m_\alpha
\overline{m}^\alpha=+1$ are the only nonvanishing products among
the tetrad vectors. Then the electromagnetic tensor
$F_{\alpha\beta}={\rm Re}({\cal F}_{\alpha\beta})$ can be written
(see, e.g., \cite{frolov} --\cite{penrose}) as
\begin{equation}
\label{p2}
{\cal F}_{\alpha\beta}=\Phi\;{l}_{[\alpha} m_{\beta]}+
\Psi\;{l}_{[\alpha} \overline{m}_{\beta]}\;,
\end{equation}
where $\Phi$ and $\Psi$ are complex scalar functions. In the rest frame of an observer with 4-velocity
$u^\alpha$ the components of the electric and magnetic field vectors are defined as
$E^\alpha=F^{\alpha\beta}u_\beta$ and $H^\alpha=(-1/2) \epsilon^{\alpha\beta\gamma\delta}F_{\gamma\delta}u_\beta$, respectively.

In what follows it is useful to introduce a local orthonormal reference frame based on a restricted set of observers that all see the electromagnetic wave traveling in
the $+z$ direction; i.e., the observers use a tetrad frame
$e^\alpha_{\;(\beta)}$ such that
\begin{equation}
\label{p3}
e^\alpha_{\;(0)}=u^\alpha\;\quad\quad,\quad\quad
e^\alpha_{\;(3)}=(-{l}_\alpha u^\alpha)^{-1}\left[{l}^\alpha+
({l}_\beta u^\beta)u^\alpha\right]\;,
\end{equation}
and $e^\alpha_{\;(1)}$, $e^\alpha_{\;(2)}$ are two unit spacelike vectors
orthogonal to each other as well as to both $e^\alpha_{\;(0)}$ and
$e^\alpha_{\;(3)}$ \cite{index} -- \cite{tetrad}. The vectors $e^\alpha_{\;(1)}$ and
$e^\alpha_{\;(2)}$ play a significant role in the discussion of polarized
radiation. In fact, the connection between the null tetrad and the frame $e^\alpha_{\;(\beta)}$ is given by
$l^\alpha=-(l_\gamma u^\gamma)(e^\alpha_{\;(0)}+e^\alpha_{\;(3)})\;$,
$n^\alpha=-\frac{1}{2}(l_\gamma u^\gamma)(e^\alpha_{\;(0)}-e^\alpha_{\;(3)})\;$,
$m^\alpha=2^{-1/2}(e^\alpha_{\;(1)}+ie^\alpha_{\;(2)})\;$ and
$\overline{m}^\alpha=2^{-1/2}(e^\alpha_{\;(1)}-ie^\alpha_{\;(2)})\;$. The
vectors $e^\alpha_{\;(1)}$ and
$e^\alpha_{\;(2)}$ are defined up to an arbitrary rotation in space; their more specific definitions will be given in section 7.

Vectors of the null tetrad $({l}^\alpha, n^\alpha,
m^\alpha, \overline{m}^\alpha)$  and those of $e^\alpha_{\;(\beta)}$
are parallel transported along the null geodesics. Thus, from the
definition (\ref{p2}) and equation (\ref{p1}) it follows that the amplitude of
the electromagnetic wave propagates according to the law
\begin{equation}
\label{p4}
{d\Phi\over d\lambda}+\theta\Phi=0\;,\qquad\qquad\quad{d\Psi\over d\lambda}+\theta\Psi=0\;.
\end{equation}
If ${\cal A}$ is the area of the cross section of a congruence of light rays, then
\begin{equation}
\label{p41}
{d{\cal A}\over d\lambda}=2\theta{\cal A}\;.
\end{equation}
Thus, ${\cal A}\;|\Phi|^2$ and ${\cal A}\;|\Psi|^2$  remain constant along the congruence of light rays $l^\alpha$.

The null tetrad frame ($l^\alpha, n^\alpha, m^\alpha, \overline{m}^\alpha$) and the associated orthonormal tetrad frame $e^\alpha_{\;(\beta)}$ are not unique. In the JWKB (or geometric optics) approximation, the null ray follows a geodesic with tangent vector $l^\alpha=dx^\alpha/d\lambda$, where $\lambda$ is an affine parameter. This parameter is defined up to a linear transformation $\lambda'=\lambda/A$+const., where $A$ is a nonzero constant, so that $l'^\alpha=dx^\alpha/d\lambda'=A\;l^\alpha$. The null tetrad is then ($l'^\alpha, n'^\alpha, m^\alpha, \overline{m}^\alpha$) with $n'^\alpha=A^{-1}n^\alpha$; this affine transformation leaves the associated tetrad $e^\alpha_{\;(\beta)}$ unchanged. Moreover, we note that at each event, the tetrad frame $e^\alpha_{\;(\beta)}$ is defined up to a Lorentz transformation. We are interested, however, in a subgroup of the Lorentz group that leaves $l^\alpha$ invariant, i.e. $\Lambda^\alpha_{\;\beta} l^\beta=l^\alpha$. This subgroup, which is the {\it little group} of the null vector $l^\alpha$, is isomorphic to the Euclidean group in the plane. This consists of translations plus a rotation. The translation part is a two-dimentional Abelian subgroup given by
\begin{eqnarray}
\label{abg1}
n'^\alpha&=&n^\alpha+B\; m^\alpha+\overline{B}\;\overline{m}^\alpha+|B|^2\;l^\alpha\;,
\\\label{abg2}
m'^\alpha&=&m^\alpha+\overline{B}\;l^\alpha\;.
\end{eqnarray}
Let us note that this transformation leaves the electromagnetic tensor $F_{\alpha\beta}$ given in (\ref{p2}) invariant; hence, this is the {\it gauge subgroup} of the {\it little group} of $l^\alpha$.

The rotation part of the subgroup under discussion is simply given by $n'^\alpha=n^\alpha$ and  $m'^\alpha=C m^\alpha$, where $C=\exp(-i\Theta)$. This corresponds to a simple rotation by a constant angle $\Theta$ in the ($e^\alpha_{\;(1)}, e^\alpha_{\;(2)}$) plane, i.e.,
\begin{eqnarray}
\label{abg3}
e'^\alpha_{(1)}&=&\cos\Theta\; e^\alpha_{\;(1)}+\sin\Theta\; e^\alpha_{\;(2)}\;,
\\\label{abg4}
e'^\alpha_{(2)}&=&-\sin\Theta\; e^\alpha_{\;(1)}+\cos\Theta\; e^\alpha_{\;(2)}\;.
\end{eqnarray}

Finally, we note that a {\it null rotation} is an element of the four-parameter group of transformations given by
\begin{eqnarray}
\label{abg5}
l'^\alpha&=&Al^\alpha\;,
\\\label{abg6}
n'^\alpha&=&A^{-1}n^\alpha+B\; m^\alpha+\overline{B}\;\overline{m}^\alpha+|B|^2\; A l^\alpha\;,
\\\label{abg7}
m'^\alpha&=&C\left(m^\alpha+\overline{B}\;Al^\alpha\right)\;.
\end{eqnarray}
A null rotation is the most general transformation of the local null tetrad frame that leaves the spatial {\it direction} of the null vector $l^\alpha$ invariant.

\section{Equations of Light Geodesics and Their Solutions}

We consider the motion of a light particle (``photon'') in the background
gravitational
field described by the metric (\ref{2a}). No back action of the
photon on the gravitational field is assumed. Hence, we are allowed to use
equations of light geodesics (\ref{ff2}) with ${l}^\alpha=dx^\alpha/d\lambda$
directly applying the metric tensor in question.
Let the motion of the photon be defined by fixing the mixed
initial-boundary conditions
\begin{equation}
\label{7a}
{\bf x}(t_{0})={\bf x}_{0}\;, \hspace{2 cm}
{\displaystyle {d{\bf x}(-\infty ) \over dt}}%
={\bf k}\;,
\end{equation}
where $|{\bf k}|^{2}=1$ and, henceforth, the spatial components of
vectors are denoted by bold letters. These conditions define the coordinates
${\bf x}_{0}$ of the photon at the moment of emission of light, $t_{0}$, and its
velocity at the infinite past and infinite distance from the origin of the
spatial coordinates (that is, at the so-called past null infinity denoted by
${\cal J}_-$ \cite{MTW}).

In the underlying inertial frame of the background flat space-time the unperturbed trajectory of the
light ray is a straight line
\begin{eqnarray}
\label{8a}
x^i(t)&=&x^i_N(t)=x^i_0+k^i\left(t-t_0\right)\;,\nonumber
\end{eqnarray}
where $t_0$, $x^i_0$, and $(k^i)={\bf k}$ have been defined in equation (\ref{7a}).
It is convenient to introduce a new independent parameter $\tau$ along the
photon's trajectory according to the rule \cite{KSGE},
\begin{eqnarray}
\label{9}
\tau&=& {\bf k}\cdot{\bf x}_N(t)=t-t_0+{\bf k}\cdot{\bf x}_{0}\;.
\end{eqnarray}
The time $t_0$ of the light signal's emission
corresponds to $\tau=\tau_0$, where
$\tau_0={\bf k}\cdot{\bf x}_{0}$, and
$\tau=0$ corresponds to the coordinate time $t=t^{\ast}$, where
\begin{eqnarray}
\label{uuu}
t^{\ast}&=&
t_0-{\bf k}\cdot{\bf x}_{0}\;.
\end{eqnarray}
This is the time of the closest approach of the unperturbed trajectory of
the photon to the origin of an asymptotically flat harmonic coordinate system.
We emphasize that the numerical value of the moment $t^{\ast}$ is constant for
a chosen trajectory of light ray and
depends only on the space-time coordinates of the point of emission
of the photon and
the point of its observation. Thus, we find the relationship
\begin{eqnarray}
\label{10a}
\tau&\equiv&t-t^{\ast}\;,
\end{eqnarray}
which reveals that the differential identity $dt=d\tau$ is valid and,
for this reason,
the integration along the light ray's path with respect to time $t$ can be
always replaced by the integration with respect to $\tau$ with the corresponding shift in the limits of integration.

Making use of the parameter $\tau$, the equation of the unperturbed trajectory
of the light ray can be represented as
\begin{eqnarray}
\label{11a}
x^i(\tau)&=&x^i_N(\tau)=k^i \tau+\xi^i\;.
\end{eqnarray}
The constant vector $(\xi^i)={\bm{\xi}}={\bf k}\times ({\bf x}_{0}
\times {\bf k})$
is called the impact parameter of the unperturbed trajectory of
the light ray with respect to the origin of the coordinates, $d=|{\bm{\xi}}|$ is the length of the impact parameter. We note that the vector ${\bm{\xi}}$ is
transverse to the vector ${\bf k}$ and
directed from the origin of the coordinate system towards the point of the
closest approach of the unperturbed path of the light ray to the origin as depicted in Figure \ref{spinfig1}.

The equations of light geodesics can be expressed in the first post-Minkowskian
approximation as follows (for more details see paper
\cite{KSGE})
\begin{eqnarray}
\label{14a}
\ddot{x}^{i}(\tau)&=&
\frac{1}{2}k_{\alpha}k_{\beta}
{\hat{\partial}}_i h^{\alpha\beta}(\tau,{\bm {\xi}})-
{\hat{\partial}}_{\tau}\left[
k_{\alpha}h^{\alpha i}(\tau,{\bm {\xi}})+\frac{1}{2}k^i
h^{00}(\tau,{\bm {\xi}})-\frac{1}{2}k^ik_p k_q
h^{pq}(\tau,{\bm {\xi}})\right]\;,
\end{eqnarray}
where an overdot denotes differentiation with respect to time,
${\hat{\partial}}_{\tau}\equiv\partial/\partial\tau$,
${\hat{\partial}}_i\equiv P_{ij}\partial/\partial \xi^j$,
$k^\alpha=(1,k^i)$, $k_\alpha=(-1,k_i)$, $k^i=k_i$,
$P_{ij}=\delta_{ij}-k_i k_j$ is the operator of projection onto the plane
orthogonal to the vector ${\bf k}$, and $h^{\alpha\beta}(\tau,{\bm {\xi}})$ is simply $h^{\alpha\beta}(t,{\bf x})$ with $t=\tau+t^*$ and ${\bf x}={\bf k}\tau+{\bm{\xi}}$. For a given null ray, all quantities on the right side
of equation (\ref{14a}) depend on the running parameter $\tau$ and
the parameter $\bm {\xi}$ which is assumed to be constant. Hence, equation
(\ref{14a}) should be considered as an ordinary second-order differential
equation in the variable $\tau$.

Perturbations of the trajectory of the photon are found by straightforward
integration of the equations of light geodesics (\ref{14a}).
Performing the calculations we find
\begin{eqnarray}
\label{25}
\dot{x}^i(\tau)&=&k^i+\dot{\Xi}^i(\tau)\;,\\
\label{26}
x^i(\tau)&=&x^i_N(\tau)+\Xi^i(\tau)-\Xi^i(\tau_0)\;,
\end{eqnarray}
where $\tau=t-t^*$ and $\tau_0=t_0-t^*$ correspond, respectively, to the moments of observation
and emission of the photon. The functions $\dot{\Xi}^i(\tau)$ and
$\Xi^i(\tau)$ are given as
follows
\begin{eqnarray}
\label{27}
\dot{\Xi}^i(\tau)&=&\frac{1}{2}k_{\alpha}k_{\beta}
{\hat{\partial}}_i B^{\alpha\beta}(\tau)-
k_{\alpha}h^{\alpha i}(\tau)-\frac{1}{2}k^i
h^{00}(\tau)+\frac{1}{2}k^ik_p k_q
h^{pq}(\tau)\;,\\\nonumber\\
\label{28}
\Xi^i(\tau)&=&\frac{1}{2}k_{\alpha}k_{\beta}
{\hat{\partial}}_i D^{\alpha\beta}(\tau)-
k_{\alpha}B^{\alpha i}(\tau)-\frac{1}{2}k^i
B^{00}(\tau)+\frac{1}{2}k^ik_p k_q
B^{pq}(\tau)\;,
\end{eqnarray}
where it is implicitly assumed that $h^{\alpha\beta}(-\infty)=0$, and the integrals $B^{\alpha\beta}(\tau)$ and $D^{\alpha\beta}(\tau)$ are given by
\begin{equation}
\label{f1}
B^{\alpha\beta}(\tau)=B^{\alpha\beta}_M(\tau)+B^{\alpha\beta}_S(\tau)\;,
\quad\quad\quad\quad B^{\alpha\beta}_M(\tau)=\int^\tau_{-\infty}
h^{\alpha\beta}_M(\sigma,{\bm {\xi}})d\sigma\;,
\quad\quad\quad\quad B^{\alpha\beta}_S(\tau)=\int^\tau_{-\infty}
h^{\alpha\beta}_S(\sigma,{\bm {\xi}})d\sigma\;,
\end{equation}
\begin{equation}
\label{f2}
D^{\alpha\beta}(\tau)=D^{\alpha\beta}_M(\tau)+D^{\alpha\beta}_S(\tau)\;,
\quad\quad\quad\quad D^{\alpha\beta}_M(\tau)=\int^\tau_{-\infty}
B^{\alpha\beta}_M(\sigma,{\bm {\xi}})d\sigma\;,
\quad\quad\quad\quad D^{\alpha\beta}_S(\tau)=\int^\tau_{-\infty}
B^{\alpha\beta}_S(\sigma,{\bm {\xi}})d\sigma\;.
\end{equation}
Integrals (\ref{f1}) and (\ref{f2}) and their derivatives are calculated in
Appendix C by extending a method developed in \cite{KSGE} and \cite{KS}. It is important to emphasize, however, that in the case the body moves along a straight line with constant velocity and spin, one can obtain the same results by direct computation without employing the methods used in Appendix C.

Equation (\ref{26}) can be used for the formulation of the boundary-value problem
for the equation of
light geodesics where the initial position,
${\bf x}_0={\bf x}(t_0)$, and final position,
${\bf x}={\bf x}(t)$, of the photon are prescribed for finding the solution of the light trajectory. This is in contrast to our original boundary-value problem, where the
initial position ${\bf x}_0$ of the photon and the direction of light
propagation ${\bf k}$ given at the past null infinity were specified. All that we
need for the solution of the new boundary-value problem is the relationship
between the unit vector ${\bf k}$ and the unit vector
\begin{eqnarray}
\label{vvv}
{\bf K}= -\;\frac{{\bf x}-{\bf x}_0}{|{\bf x}-{\bf x}_0|}\;,
\end{eqnarray}
which defines a geometric (coordinate) direction
of the light propagation from the observer to the source of light as if
the space-time were flat. The formulas (\ref{26}) and
(\ref{28}) yield
\begin{eqnarray}
\label{29}
k^i&=&-K^i-\beta^i(\tau,{\bm{\xi}})+\beta^i(\tau_0,{\bm{\xi}})\;,
\end{eqnarray}
where the relativistic corrections to the vector $K^i$ are given by
\begin{eqnarray}
\label{30}
\beta^i(\tau,{\bm{\xi}})&=&\frac{\frac{1}{2}k_{\alpha}k_{\beta}
{\hat{\partial}}_i D^{\alpha\beta}(\tau)-
k_{\alpha}P_{ij}B^{\alpha j}(\tau)}{|{\bf x}(\tau)-{\bf x}_0|}\;.
\end{eqnarray}
We emphasize that the vectors $(\beta^i)\equiv{\bm{\beta}}(\tau,{\bm{\xi}})$ and
$(\beta^i_0)\equiv{\bm{\beta}}(\tau_0,{\bm{\xi}})$ are orthogonal to
${\bf k}$ and are evaluated at the points of observation and emission of the
photon, respectively.
The relationships obtained in this section are used for the discussion of
observable
relativistic effects in the following sections.

\section{Gravitomagnetic Effects in Pulsar Timing, Astrometry and Doppler Tracking}

\subsection{Shapiro Time Delay in Binary Pulsars}

We shall give in this paragraph the relativistic time delay
formula for the case of the propagation of light through the {\it
non-stationary} gravitational field of arbitrary-moving and
rotating body. The total time of propagation of an electromagnetic
signal from the point ${\bf x}_0$ to the point ${\bf x}$ is
derived from equations (\ref{26}) and (\ref{28}). First, we use 
equation (\ref{26}) to express the difference ${\bf x}-{\bf x}_0$
via the other terms of this equation. Then, we find the total coordinate time of propagation
of light, $t-t_0$ from
\begin{eqnarray}
\label{qer}
t-t_0&=&|{\bf x}-{\bf x}_0|+\Delta_M(t,t_0)+\Delta_S(t,t_0)\;,
\end{eqnarray}
where $|{\bf x}-{\bf x}_0|$ is the usual Euclidean distance
between the points of emission, ${\bf x}_0$, and observation,
${\bf x}$, of the photon, $\Delta_M(t,t_0)$ is the Shapiro time
delay produced by the gravitoelectric field of a point-like
massive body, and $\Delta_S(t,t_0)$ is the Shapiro time delay
produced by the gravitomagnetic field of the spinning source. The
term $\Delta_M(t,t_0)$ is discussed in detail in \cite{kop2001}. The
new term $\Delta_S(t,t_0)$ is given by (cf. equation (\ref{f5a})
of Appendix C)
\begin{eqnarray}
\label{shapd}
\Delta_S(t,t_0)=B_S(\tau)-B_S(\tau_0)\;,
\end{eqnarray}
\begin{eqnarray}
\label{ur}
B_S(\tau)&\equiv&\frac{1}{2}k_{\alpha}k_{\beta}B^{\alpha\beta}_S(\tau)=
2\;\frac{1-{\bf k}\cdot{\bf v}}{\sqrt{1-v^2}}\frac{k_\alpha r_\beta
S^{\alpha\beta}}{(r-{\bf v}\cdot{\bf r})(r-{\bf k}\cdot{\bf r})}\;,
\end{eqnarray}
\begin{eqnarray}
\label{ur1}
B_S(\tau_0)&\equiv&\frac{1}{2}k_{\alpha}k_{\beta}B^{\alpha\beta}_S(\tau_0)=
2\;\frac{1-{\bf k}\cdot{\bf v}_0}{\sqrt{1-v_0^2}}\frac{k_\alpha r_{0\beta}
S^{\alpha\beta}_0}{(r_0-{\bf v}_0\cdot{\bf r}_0)(r_0-{\bf k}\cdot{\bf r}_0)}\;,
\end{eqnarray}
where the times $\tau=t-t^\ast$ and $\tau_0=t_0-t^\ast$ are
related to the retarded times $s$ and $s_0$ via equation
(\ref{fgh2}), ${\bf r}={\bf x}-{\bf z}$, ${\bf r}_0= {\bf
x}_0-{\bf z}_0$, $r_0=|{\bf r}_0|$, ${\bf x}={\bf x}(t)$, ${\bf
x}_0={\bf x}(t_0)$, ${\bf z}={\bf z}(s)$, ${\bf z}_0={\bf
z}(s_0)$, ${\bf v}={\bf v}(s)$, and ${\bf v}_0={\bf v}(s_0)$. It
is worthwhile to note that in the approximation where one can
neglect relativistic terms in the relationship (\ref{29}) between
vectors ${\bf k}$ and $-{\bf K}$, the expression for ${\bf r}$ is
given by
\begin{eqnarray}
\label{xo}
{\bf r}&=&D{\bf k}+{\bf x}_0-{\bf z}(s)\;,
\end{eqnarray}
where $D=|{\bf x}-{\bf x}_0|$. In the case of a binary pulsar,
the distance $D$ between the pulsar and the solar system is much
larger than that between the point of the emission of the radio
pulse, ${\bf x}_0$ and the pulsar or its companion ${\bf z}(s)$.
Hence,
\begin{eqnarray}
\label{xo1} {\bf r}-{\bf k}r&=&-\frac{[{\bf k}\times({\bf
x}_0-{\bf z})]^2}{2D}\;{\bf k}\;,
\end{eqnarray}
where terms of higher order in the ratio $|{\bf x}_0-{\bf z}|/D$ are
neglected.

The result (\ref{shapd}) can be used, for example, to find out the
spin-dependent relativistic correction $\Delta_S$ to the timing
formula of binary pulsars. The binary pulsar consists of two
bodies --- the pulsar itself (index ``p") and its companion (index
``c"); in what follows we let ${\bf r}_p={\bf x}-{\bf z}_p(s)$,
${\bf r}_c={\bf x}-{\bf z}_c(s)$, ${\bf r}_{0p}={\bf x}_0-{\bf
z}_p(s_0)$, and ${\bf r}_{0c}={\bf x}_0- {\bf z}_c(s_0)$, where
${\bf z}_p$ and ${\bf z}_c$ are coordinates of the pulsar and its
companion, respectively. According to \cite{KS}, the difference
between the instants of time $s$ and $s_0$, $s$ and $t_0$, and
$s_0$ and $t_0$ for binary pulsars is of the order of the time
required for light to cross the system, that is of the order of a
few seconds. Thus, we can expand all quantities depending on time
in the neighborhood of the instant $t_0$ and make use of the
approximations ${\bf r}_p={\bf x}-{\bf
z}_p(t_0)\equiv{\bm{\rho}}_p$, ${\bf r}_c={\bf x}-{\bf
z}_c(t_0)\equiv{\bm{\rho}}_c$, ${\bf r}_{0p}={\bf x}_0-{\bf
z}_p(t_0)\equiv{\bm{\rho}}_{0p}$, ${\bf r}_{0c}={\bf x}_0-{\bf
z}_c(t_0)\equiv{\bm{\rho}}_{0c}$, ${\bm{\cal J}_p}={\bm{\cal
J}_{0p}}$, and ${\bm{\cal J}_c}={\bm{\cal J}_{0c}}$. The next
approximation used in the calculations is
\begin{eqnarray}
\label{ap1}
{\bf x}_0&=&{\bf z}_{p}(t_0)+{\bf k}\;X\;,
\end{eqnarray}
where $X$ is the distance from the pulsar's center of mass to the
point of emission of radio pulses \cite{10}. One can also see that
${\bm{\rho}}_{0c}={\bf R}+{\bf k}X$, where ${\bf R}={\bf
z}_p(t_0)- {\bf z}_c(t_0)$, i.e. the radius-vector of the pulsar
with respect to the companion. Taking these approximations into account, the
following equalities hold
\begin{eqnarray}
\label{xo3}
{\bm{\rho}_c}-{\bf k}{\rho}_c&=&
-\frac{({\bf k}\times{\bf R})^2}{2D}{\bf
k}\;,
\end{eqnarray}
\begin{eqnarray}
\label{xo4}
{\bm{\rho}_p}-{\bf k}{\rho}_p&=&0\;.
\end{eqnarray}
Moreover,
\begin{eqnarray}
\label{ar3}
{\rho}_c-{\bf k}\cdot{\bm{\rho}}_c&=&\frac{{\rho}_{0c}^2-({\bf
k}\cdot{\bm{\rho}}_{0c})^2}{2{\rho}_c}=
\frac{R^2-({\bf k}\cdot{\bf R})^2}{2{\rho}_c} \;.
\end{eqnarray}

The primary contribution to the function $\Delta_S(t,t_0)$ is
obtained after expansion of the expressions for $B(\tau)$ and
$B(\tau_0)$ in powers of $v/c$ and picking up all
velocity-independent terms. Accounting for equations (\ref{xo3}) -
(\ref{ar3}) and (\ref{f9}) this procedure gives the time delay
correction $\Delta_S$ to the standard timing formula as follows
\begin{eqnarray}
\label{bubu} \Delta_S&=&\frac{2{\bm{\cal J}_p}\cdot({\bf k}
\times{\bf r}_p)}{r_p(r_p-{\bf k}\cdot{\bf r}_p)}
+\frac{2{\bm{\cal J}_c}\cdot({\bf k} \times{\bf
r}_c)}{r_c(r_c-{\bf k}\cdot{\bf r}_c)} -\frac{2{\bm{\cal
J}_{0p}}\cdot({\bf k} \times{\bf r}_{0p})}{r_{0p}(r_{0p}-{\bf k}
\cdot{\bf r}_{0p})}-\frac{2{\bm{\cal J}_{0c}}\cdot({\bf k}
\times{\bf r}_{0c})}{r_{0c}(r_{0c}-{\bf k} \cdot{\bf r}_{0c})}\;.
\end{eqnarray}
By means of (\ref{ap1}) and (\ref{xo4}) we find that the first and
third terms in (\ref{bubu}) drop out. Neglecting terms of order
$X/R$ we can see that ${\bm{\rho}}_{0c}={\bf R}$, which
brings $\Delta_S$ into the form
\begin{eqnarray}
\label{ar2}
\Delta_S&=&2{\bm{\cal J}_{0c}}\cdot({\bf k}
\times{\bf R})\left[\frac{1}{{\rho}_c({\rho}_c-{\bf k}\cdot{\bm{\rho}}_c)}
-\frac{1}{R(R-{\bf k}\cdot{\bf R})}\right]\;.
\end{eqnarray}
Making use of (\ref{ar3}) transforms expression (\ref{ar2}) to
the simpler form
\begin{eqnarray}
\label{ar4}
\Delta_S&=&\frac{2{\bm{\cal J}_{0c}}\cdot({\bf k}
\times{\bf R})}{R(R+{\bf k}\cdot{\bf R})}=
-\frac{2{\bm{\cal J}_{0c}}\cdot({\bf K}
\times{\bf R})}{R(R-{\bf K}\cdot{\bf R})}\;,
\end{eqnarray}
where the unit vector ${\bf K}$ is defined in (\ref{vvv}).

Formula (\ref{ar4}) coincides exactly with that  obtained on the
basis of the post-Newtonian expansion of the metric tensor and
subsequent integration of light-ray propagation in the static
gravitational field of the pulsar companion \cite{11}. In
principle, the additional time delay caused by the spin of the
companion might be used for testing whether the companion is a
black hole or not \cite{12}. However, as shown in
\cite{11}, the time delay due to the spin is not separable from
the delay caused by the bending of light rays in the gravitational
field of the companion \cite{13}. For this reason, the delay
caused by the spin is not a directly measurable quantity and can
not be effectively used for testing the presence of the black hole
companion of the pulsar \cite{11}.

\subsection{Deflection of Light in Gravitational Lenses and by the Solar System}

Let us assume that the observer is at rest at an event with space-time
coordinates $(t,{\bf x})$. The observed direction ${\bf s}$
to the source of light has
been derived in \cite{KSGE} and is given by
\begin{eqnarray}
\label{ar5}
{\bf s}&=&{\bf K}+{\bm{\alpha}}+{\bm{\beta}}-{\bm{\beta}}_0+{\bm{\gamma}}\;,
\end{eqnarray}
where the unit vector ${\bf K}$ is given in (\ref{vvv}), the relativistic corrections
${\bm{\beta}}$ and ${\bm{\beta}}_0$ are defined in (\ref{30}), 
${\bm{\alpha}}$ describes the overall effect of deflection of the light-ray
trajectory in the plane of the sky, and ${\bm{\gamma}}$ is related to the distortion of the local
coordinate system of the observer with respect to the underlying global coordinate system
used for the calculation of the propagation of light rays.

More precisely, the quantity ${\bm{\alpha}}$ can be expressed as
\begin{equation}
\label{ar6}
\alpha^i=\alpha^i_M+\alpha^i_S\;,
\quad\quad\quad\quad\alpha^i_M=
-\hat{\partial}_i B_M(\tau)+k_\alpha P^i_{\;j} h^{\alpha j}_M(\tau)\;,
\quad\quad\quad\quad\alpha^i_S=
-\hat{\partial}_i B_S(\tau)+k_\alpha P^i_{\;j} h^{\alpha j}_S(\tau)\;,
\end{equation}
the quantities ${\bm{\beta}}={\bm{\beta}}_M+{\bm{\beta}}_S$ and
${\bm{\beta}}_0={\bm{\beta}}_{0M}+{\bm{\beta}}_{0S}$
are defined by (\ref{30}), and
\begin{equation}
\label{oop}
\gamma^i=\gamma^i_M+\gamma^i_S\;,
\quad\quad\quad\quad
\gamma^i_M=-\frac{1}{2}k_n P^i_{\;j} h^{nj}_M(\tau)\;,
\quad\quad\quad\quad
\gamma^i_S=-\frac{1}{2}k_n P^i_{\;j} h^{nj}_S(\tau)\;.
\end{equation}
In what follows, we neglect all terms depending on the acceleration of the
light-ray-deflecting body and the time derivative of its spin.
Light-ray deflections represented by $\alpha^i_M$, $\beta^i_M$, and $\gamma^i_M$ caused by
the mass-monopole part of the stress-energy tensor (\ref{1}) have been
calculated in \cite{KS} and will not be given here as our primary interest
in the present paper is the description of the spin-dependent gravitomagnetic
effects. Using equation
(\ref{ur}) for the function $B_S(\tau)$ and formula (\ref{poo2}),
we obtain
\begin{eqnarray}
\label{uro}
\hat{\partial}_i B_S(\tau)&=&-2\;\frac{1-{\bf k}\cdot{\bf v}}{\sqrt{1-v^2}}\left[
\frac{(1-v^2)k_\alpha r_\beta S^{\alpha\beta}P_{ij}r^j}
{(r-{\bf v}\cdot{\bf r})^3(r-{\bf k}\cdot{\bf r})}+
\frac{(1-{\bf k}\cdot{\bf v})k_\alpha r_\beta S^{\alpha\beta}P_{ij}r^j}
{(r-{\bf v}\cdot{\bf r})^2(r-{\bf k}\cdot{\bf r})^2}\right.
\\\nonumber\\\nonumber&-&\left.
\frac{k_\alpha r_\beta S^{\alpha\beta}P_{ij}v^j}
{(r-{\bf v}\cdot{\bf r})^2(r-{\bf k}\cdot{\bf r})}-
\frac{P_{ij}k_\alpha S^{\alpha j}}
{(r-{\bf v}\cdot{\bf r})(r-{\bf k}\cdot{\bf r})}\right]\;.
\end{eqnarray}
In addition, making use of formula (\ref{f6}) yields
\begin{eqnarray}
\label{wer}
\frac{1}{2}k_\alpha k_\beta{\hat\partial}_i D^{\alpha\beta}_S(\tau)&=&
-2\;\frac{1-{\bf k}\cdot{\bf v}}{\sqrt{1-v^2}} \frac{\;P_{ij}r^j}
{r-{\bf v}\cdot{\bf r}}\;\frac{k_\alpha r_{\beta} S^{\alpha\beta}}
{(r-{\bf k}\cdot{\bf r})^2}+{2\over\sqrt{1-v^2}}{k_\alpha P_{ij}S^{\alpha j}\over r-{\bf k}\cdot{\bf r}}\;,
\end{eqnarray}
\begin{eqnarray}
\label{wera}
\frac{1}{2}k_\alpha k_\beta{\hat\partial}_i D^{\alpha\beta}_S(\tau_0)&=&
-2\;\frac{1-{\bf k}\cdot{\bf v}_0}{\sqrt{1-v^2_0}} \frac{\;P_{ij}r^j_0}
{r_0-{\bf v}_0\cdot{\bf r}_0}\;\frac{k_\alpha r_{0\beta} S^{\alpha\beta}_0}
{(r_0-{\bf k}\cdot{\bf r}_0)^2}+{2\over\sqrt{1-v_0^2}}{k_\alpha P_{ij}S_0^{\alpha j}\over r_0-{\bf k}\cdot{\bf r}_0}\;.
\end{eqnarray}
The light deflection vector, i.e. the vector connecting the undeflected image to the deflected image of the source of light, in the plane of the sky is defined in \cite{KS} and for the purely spin-induced part can be calculated from the third relation in equation (\ref{ar6}). In that relation the 
the first term is given in (\ref{uro}) and the second one can be
calculated from equation (\ref{fgh1}), which results in
\begin{eqnarray}
P_{ij}k_\alpha h^{\alpha j}_S&=&-2\sqrt{1-v^2}\left[(1-{\bf k}\cdot{\bf v})
{P_{ij}r_\alpha S^{\alpha j}\over (r-{\bf v}\cdot{\bf r})^3}-
P_{ij}v^j\;
{k_\alpha r_\beta S^{\alpha\beta}\over (r-{\bf v}\cdot{\bf r})^3}\right]\;.
\end{eqnarray}

The most interesting physical application of the formalism given in the present
section is the gravitomagnetic deflection of light in gravitational lenses
and by bodies in the solar system. In both cases the impact parameter $d$ of the
light ray is considered to be extremely small in comparison with the distances
from the body to the
observer and the source of light.
The gravitational lens approximation allows us to use the following
relationships (for more details see section 7B in \cite{KS})
\begin{equation}
{\bf r}-{\bf k} r=\bm{\zeta}\;,\quad\quad\quad\quad
{\bf r}_0+{\bf k} r_0=\bm{\zeta}_0\;,
\end{equation}
where $\zeta^i=P^i_{\;j}(x^j-z^j(s))$ and
$\zeta^i_0=P^i_{\;j}(x^j_0-z^j(s_0))$ are impact parameters
of the light ray with respect to the light-deflecting body evaluated at the moments
of observation and emission of light. Assuming that $|\bm{\zeta}|
\equiv d\ll \min[r,r_0]$ and $|\bm{\zeta}_0|\equiv d_0\ll \min[r,r_0]$,
where
$r=|{\bf x}-{\bf z}(s)|$ and $r_0=|{\bf x}_0-{\bf z}(s_0)|$ are distances
from the body to observer and to the source of light,
respectively, one can derive the following approximations
\begin{equation}
\label{g1}
r-{\bf k}\cdot{\bf r}={d^2\over 2r}\;,\quad\quad\quad\quad
r_0-{\bf k}\cdot{\bf r}_0=2r_0\;,
\end{equation}
along with
\begin{equation}
\label{g2}
r-{\bf v}\cdot{\bf r}=r(1-{\bf k}\cdot{\bf v})\;,\quad\quad\quad\quad
r_0-{\bf v}_0\cdot{\bf r}_0=r_0(1+{\bf k}\cdot{\bf v}_0)\;.
\end{equation}
Hence, the relativistic spin-induced deflection of light $\alpha^i_S$ in (\ref{ar6}) can be calculated from
\begin{eqnarray}
\label{pppo}
\alpha^i_S&=&{4\over\sqrt{1-v^2}}\left[{2\;k_\alpha r_\beta S^{\alpha\beta}\zeta^i\over d^4}-{\;k_\alpha P_{ij} S^{\alpha j}\over d^2}\right]\;,
\end{eqnarray}
where we have neglected all residual terms of $O(d/r)$. After substituting the expressions given in (\ref{f8}) and (\ref{f9}) in (\ref{pppo}), one obtains the analytic representation of the relativistic deflection of light valid for the body having arbitrary high speed ${\bf v}$ and constant spin ${\bm{\cal J}}$
\begin{eqnarray}
\label{pp11}
\alpha^i_S&=&{8\gamma^2\over d^4}\left[{\bm{\cal J}}\cdot({\bf k}\times{\bm\zeta})+{\bm{\cal J}}\cdot({\bm\zeta}\times{\bf v})+{1-\gamma\over \gamma v^2}({\bf v}\cdot{\bm{\cal J}})({\bf k}\times {\bm\zeta})\cdot {\bf v}  \right]\;\zeta^i\nonumber\\\nonumber\\&&\hspace{2cm}-{4\gamma^2\over d^2}\left[ P_{ij}({\bf v}\times{\bm{\cal J}})^j-({\bf k}\times{\bm{\cal J}})^i-{1-\gamma\over \gamma v^2}({\bf v}\cdot{\bm{\cal J}})({\bf k}\times{\bf v})^i \right]\;.
\end{eqnarray}
In the case of slow motion, the Taylor expansion of (\ref{pp11}) with respect to the parameter $v/c$ yields
\begin{eqnarray}
\label{pp00}
\alpha^i_S&=&{8{\bm{\cal J}}\cdot({\bf k}\times{\bm\zeta})\zeta^i\over d^4}+{4({\bf k}\times{\bm{\cal J}})^i\over d^2}\;,
\end{eqnarray}
which exactly coincides with the previously known formula for a stationary rotating gravitational lens (see, e.g., the leading terms of equation (6.28) in \cite{KK}) derived using a different mathematical technique. One can also recast (\ref{pp00}) in a simpler gradient form
\begin{equation}
\label{poz}
\alpha^i_S=4{\partial\psi_S\over\partial\zeta^i}\;,\qquad\qquad \psi_S=4({\bf k}\times{\bm{\cal J}})^j{\partial \ln d/\partial\zeta^j}\;,
\end{equation}
where $\psi_S$ is a gravitomagnetic component of the gravitational lens potential (cf. the second term of equation (153) in \cite{KS}).

\subsection{Gravitational Shift of Frequency and Doppler Tracking}

The special relativistic treatment of the Doppler frequency shift in an inertial system of Cartesian coordinates is 
well-known. It is based on two facts: proper time runs differently for identical clocks moving with different
speeds and electromagnetic waves propagate along straight lines in such an inertial system in flat space-time (see \cite{Ocun} and \cite{KO99} for more details).
The general relativistic formulation of the frequency shift in curved space-time
is more involved.

Two definitions of the Doppler shift are used \cite{Synge} ---
in terms of energy ($A$) and in terms of frequency ($B$)
\begin{equation}
\label{d1}
(A)\quad\quad\frac{\nu }{\nu _{0}}=\frac{u^{\alpha }{l}_{\alpha }}{u_{0}^{\alpha }%
{l}_{0\alpha }}\;,\quad\quad\quad\quad\quad\quad\quad (B)\quad\quad
\frac{\nu }{\nu _{0}}=\frac{dT_{0}}{dT}\;,
\end{equation}
where $\nu_0$ and $\nu$ are emitted and observed electromagnetic
frequencies of light; here ($T_0$, $u_0^{\alpha }$)
and ($T$, $u^{\alpha }$) are respectively the proper time and 4-velocity of the source of light
and observer, and 
${l}_{0\alpha}$ and ${l}_{\alpha}$ are null 4-vectors of the light
ray at the points of emission and observation, respectively. Despite the
apparent difference in the two definitions, they are identical as equalities
$u^{\alpha}=dx^{\alpha}/dT$ and ${l}_{\alpha}=
\partial\varphi/\partial x^{\alpha}$ hold. 
In order to connect various physical
quantities at the points of emission and observation of light one has to integrate the equations of light propagation.

The integration of null geodesic equation in the case of space-times
possessing symmetries has been known for a long time and extensively
used in astronomical practice (see, e.g., \cite{K90} and references therein).
However, interesting astronomical phenomena in the propagation
of light rays in curved space-time are also caused by small time-dependent
perturbations of the background geometry. Usually, the first post-Newtonian
approximation
in the relativistic N-body problem with fixed or uniformly moving
bodies has been applied in order to consider the effects of the N-body system on
electromagnetic signals \cite{KK}.
Unfortunately, this
approximation works properly if, and only if, the time of propagation of
light is much
shorter than the characteristic Keplerian time of the N-body problem.
An adequate treatment of the effects in the
propagation of light must account for the retardation in the
propagation of gravitational field from the light-deflecting body
to the point of interaction of the field with the electromagnetic signal.

A theory of light propagation and Doppler shift that takes account of such retardation effects has been
constructed in the first post-Minkowskian approximation by Kopeikin and
Sch\"afer \cite{KS} and Kopeikin \cite{K99a}. For the sake of brevity, we do not reproduce the formalism here and restrict ourselves to the consideration of the case of
gravitational lens only. Some details of this approximation have been
given in the previous section. Making use of either of the definitions
(\ref{d1}), we obtain for the gravitational shift of frequency
\begin{eqnarray}
\label{d2}
\left(\frac{\delta \nu }{\nu _{0}}\right) _{gr}&=&
\left(-{\bf v}+{r_0\over D}{\bm{\upsilon}}+{r\over
D}{\bm{\upsilon}}_0\right)\cdot\left({\bm{\alpha}}_M+{\bm{\alpha}}_S\right)\;,
\end{eqnarray}
where $D=|{\bf x}-{\bf x}_0|$ is the distance between the source of light
and observer, $r=|{\bf x}-{\bf z}(s)|$ is the distance between the lens and
observer, $r_0=|{\bf x}_0-{\bf z}(s_0)|$ is the distance between the lens and
the source of light, ${\bm{\upsilon}}=d{\bf x}(t)/dt$ is the velocity of the source of light, ${\bm{\upsilon}}_0=d{\bf x}_0(t_0)/dt_0$ is the velocity of the observer, and $\bm{\alpha}_M$ and $\bm{\alpha}_S$ represent vectors of
the deflection of light by the gravitational lens given in (\ref{ar6}).
Formula (\ref{d2}) can be applied to the processing of Doppler
tracking data from spacecrafts in deep space. In the case of
superior conjunction of such a spacecraft with the Sun or a planet, the result
shown in (\ref{d2}) should be doubled since the light passes the light-deflecting
body twice --- the first time on its way from the emitter to the spacecraft and
the second time on its way back to the receiver. The Doppler tracking formula
after subtracting the special relativistic corrections (for details see
\cite {KS} and \cite{K99a}) assumes the following form
\begin{eqnarray}
\label{d3}
\left(\frac{\delta \nu }{\nu _{0}}\right) _{gr}&=&
\left({8GM\over c^3 d^2}+
{16G({\bf k}\times{\bm{\cal J}})\cdot{\bm{\zeta}}\over c^4 d^4}\right)
\left({\bf v}-{r_0\over D}{\bm{\upsilon}}-{r\over
D}{\bm{\upsilon}}_0\right)\cdot{\bm{\zeta}}
-{8G\over c^4 d^2}
\left({\bf v}-{r_0\over D}{\bm{\upsilon}}-{r\over
D}{\bm{\upsilon}}_0\right)\cdot({\bf k}\times{\bm{\cal J}})\;.
\end{eqnarray}
Only terms proportional to the mass $M$ of the deflector were known previously (see, e.g., \cite{KS} and \cite{bert}). With the mathematical techniques of the present paper general relativistic corrections due to the intrinsic rotation of the gravitational lens can now be calculated. For the light ray grazing the limb of the light-ray-deflecting body, the gravitomagnetic Doppler shift due to the body's rotation is smaller than the effect produced by the body's mass by terms of order $\omega_{\rm rot}L/c$, where $\omega_{\rm rot}$ is the body's angular frequency and $L$ is its characteristic radius. In the case of the Sun, the effect reaches a magnitude of about $0.8\times 10^{-14}$, which is measurable in practice taking into account the current stability and accuracy of atomic time and frequency standards ($\sim 10^{-16}$, cf. \cite{atom} and \cite{sll}). For Jupiter, the corresponding estimate of the gravitomagnetic Doppler shift is $0.7\times 10^{-15}$, which is also, in principle, measurable.

\section{The Skrotskii Effect For Arbitrary Moving Pole-Dipole Massive Bodies }

\subsection{Relativistic Description of Polarized Radiation}

The polarization properties of electromagnetic radiation are defined in terms of the electric field measured by an observer. Let us start with a general electromagnetic radiation field $F_{\alpha\beta}$
and define the complex field ${\cal F}_{\alpha\beta}$ such that $F_{\alpha\beta}= {\rm Re}({\cal F}_{\alpha\beta})$ and $E_\alpha={\rm Re}({\cal E}_\alpha)$, where ${\cal E}_\alpha={\cal F}_{\alpha\beta}u^\beta$ is the complex electric field.
In the rest frame of an observer with 4-velocity $u^\alpha$,
the intensity and polarization properties of the radiation are describable in terms of
the tensor
\begin{equation}
\label{z2}
J_{\alpha\beta}=
<{\cal E}_\alpha \overline{\cal E}_\beta>\;,
\end{equation}
where the angular brackets represent an ensemble average and $J_{\alpha\beta}u^\beta=0$.
The electromagnetic Stokes parameters are defined with respect to two of the four vectors of the
tetrad $e^\alpha_{\;(\beta)}$, introduced in (\ref{p3}), as follows (cf. \cite{landau} and \cite{anil})
\begin{eqnarray}
\label{z4}
S_0&=&J_{\alpha\beta}\left[e^\alpha_{\;(1)}e^\beta_{\;(1)}+
e^\alpha_{\;(2)}e^\beta_{\;(2)}\right]\;,\\
\label{z5}
S_1&=&J_{\alpha\beta}\left[e^\alpha_{\;(1)}e^\beta_{\;(1)}-
e^\alpha_{\;(2)}e^\beta_{\;(2)}\right]\;,\\
\label{z6}
S_2&=&J_{\alpha\beta}\left[e^\alpha_{\;(1)}e^\beta_{\;(2)}+
e^\alpha_{\;(2)}e^\beta_{\;(1)}\right]\;,\\
\label{z7}
S_3&=&iJ_{\alpha\beta}\left[e^\alpha_{\;(1)}e^\beta_{\;(2)}-
e^\alpha_{\;(2)}e^\beta_{\;(1)}\right]\;.
\end{eqnarray}
Using equation (\ref{z2}), the Stokes parameters can be expressed in the standard way \cite{landau} in a linear polarization basis as 
\begin{eqnarray}
\label{z8}
S_0&=&<|{\cal E}_{(1)}|^2+|{\cal E}_{(2)}|^2>\;,\\
\label{z9}
S_1&=&<|{\cal E}_{(1)}|^2-|{\cal E}_{(2)}|^2>\;,\\
\label{z10}
S_2&=&<{\cal E}_{(1)}\overline{\cal E}_{(2)}+\overline{\cal E}_{(1)}{\cal E}_{(2)}>\;,\\
\label{z11}
S_3&=&i<{\cal E}_{(1)}\overline{\cal E}_{(2)}-\overline{\cal E}_{(1)}{\cal E}_{(2)}>\;,
\end{eqnarray}
where ${\cal E}_{(n)}={\cal E}_\alpha e^\alpha_{\;(n)}$ for $n=1,2$.    Under the gauge subgroup of the little group of $l^\alpha$, the Stokes parameters remain invariant. However, for a constant rotation of angle $\Theta$ in the ($e^\alpha_{\;(1)}, e^\alpha_{\;(2)}$) plane, $S'_0=S_0$, $S'_1=S_1\cos 2\Theta+S_2\sin 2\Theta$, $S'_2=-S_1\sin 2\Theta+S_2\cos 2\Theta$, and $S'_3=S_3$. This is what would be expected for a spin-1 field. That is, under a duality rotation of $\Theta=\pi/2$, one linear polarization state turns into the other.

For the null field (\ref{p2}) under consideration in this paper,
\begin{eqnarray}
\label{wzp}
{\cal E}_\alpha&=&{1\over 2}\omega(\Phi\;m_\alpha+\Psi\;\overline{m}_\alpha)\;,
\end{eqnarray}
where $\omega=-l^\alpha u_\alpha$ is the {\it constant} frequency of the ray measured by the observer with 4-velocity $u^\alpha$. Changing from the circular polarization basis (\ref{wzp}) with amplitudes $\omega\Phi/2$ and $\omega\Psi/2$ to the linear polarization basis ${\cal E}^\alpha={\cal E}_{(1)}e^\alpha_{\;(1)}+{\cal E}_{(2)}e^\alpha_{\;(2)}$, we find that ${\cal E}_{(1)}=\omega(\Phi+\Psi)/\sqrt{8}$ and ${\cal E}_{(2)}=i\omega(\Phi-\Psi)/\sqrt{8}$. The variation of the Stokes parameters along the ray are essentially given by equation (\ref{p4}), since the frequency $\omega$ is simply a constant parameter along the ray given by $\omega=dt/d\lambda=d\tau/d\lambda$.

The polarization vector ${\bf P}$ and the degree of polarization $P=|{\bf P}|$ can be defined in terms of the Stokes parameters $(S_0,{\bf S})$ by ${\bf P}={\bf S}/S_0$ and $P=|{\bf S}|/S_0$, respectively. Any partially polarized wave may be thought of as an incoherent superposition of a completely polarized wave with Stokes parameters $(PS_0, {\bf S})$ and a completely unpolarized wave with Stokes parameters $(S_0-PS_0, {\bf 0})$, so that $(S_0, {\bf S})=(PS_0, {\bf S})+(S_0-PS_0, {\bf 0})$. For completely polarized waves, ${\bf P}$ describes the surface of the unit sphere introduced by Poincar\'e. The center of Poincar\'e sphere corresponds to unpolarized radiation and the interior to partially polarized radiation. Orthogonally polarized waves represent any two conjugate points on the Poincar\'e sphere; in particular, $P_1=\pm 1$ and $P_3=\pm 1$ represent orthogonally polarized waves corresponding to the linear and circular polarization bases, respectively.

Any stationary or time-dependent axisymmetric gravitational field in general causes a relativistic effect of the rotation of the polarization plane of electromagnetic waves. This effect was first discussed by Skrotskii (\cite{Sk57} 
and later by many other researches (see, e.g., \cite{mashhoon75} and \cite{kl11} and references therein). We generalize the results of previous authors to the case of spinning bodies that can move arbitrarily fast.

\subsection{Reference Tetrad Field}

The rotation of the polarization plane is
not conceivable without an unambigious definition of a local reference frame
(tetrad) constructed along the null geodesic. The null
tetrad frame based on four null vectors $(l^\alpha, n^\alpha, m^\alpha,
\overline{m}^\alpha)$ introduced in section 4 is a particular choice. As discussed in section 4, this null tetrad is intimately connected with the local frame
$(e^\alpha_{\;(0)}, e^\alpha_{\;(1)} , e^\alpha_{\;(2)} , e^\alpha_{\;(3)})$, where
the vectors $e^\alpha_{\;(0)}$ and $e^\alpha_{\;(3)}$ are defined in (\ref{p3})
and the spacelike vectors $e^\alpha_{\;(1)}$ and $e^\alpha_{\;(2)}$ are
directly related to the polarization of the electromagnetic wave.
Each vector of the tetrad $(e^\alpha_{\;(0)}, e^\alpha_{\;(1)} , e^\alpha_{\;(2)} , e^\alpha_{\;(3)})$ depends upon time and is parallel transported along the null geodesic defined by its tangent vector $l^\alpha$. To characterize the variation of $e^\alpha_{\;(\beta)}$ along the ray, it is necessary to have access to a fiducial field of tetrad frames for reference purposes. To this end, let us choose the reference frame based on the set of static observers in the background space-time.
Then, at the past null infinity, where according to our assumption the space-time is asympotically flat, one has
\begin{eqnarray}\label{dqx}
e^\alpha_{(0)}(-\infty)=(1,0,0,0)\;,\quad
e^\alpha_{(1)}(-\infty)=(0,a^1,a^2,a^3)\;,\quad
e^\alpha_{(2)}(-\infty)=(0,b^1,b^2,b^3)\;,\quad
e^\alpha_{(3)}(-\infty)=(0,k^1,k^2,k^3)\;.
\end{eqnarray}
Here the spatial vectors ${\bf a}=(a^1,a^2,a^3)$, ${\bf b}=(b^1,b^2,b^3)$, and
${\bf k}=(k^1,k^2,k^3)$ are orthonormal in the Euclidean sense and the vector ${\bf k}$ defines the spatial direction of propagation of the light ray (see equation (\ref{7a})). At the same time, the four vectors (\ref{dqx}) also form
a basis ${\bm{\epsilon}}_{(\beta)}={\epsilon}^\alpha_{(\beta)}\partial/\partial x^\alpha$
of the global harmonic coordinate system at each point of space-time,
where ${\epsilon}^\alpha_{\;(\beta)}=e^\alpha_{\;(\beta)}(-\infty)$.
However, it is worth noting that this coordinate basis is not orthonormal at an arbitrary point in space-time. Nevertheless, it is possible to construct an orthonormal basis $\omega^\alpha_{\;(\beta)}$ at each point by making use of a linear transformation
$\omega^{\alpha}_{\;(\beta)}=\Lambda^\alpha_{\;\gamma}{\epsilon}^\gamma_{\;(\beta)}$
such that the transformation matrix is given in the linear approximation by
\begin{equation}
\Lambda^0_{\;0}=1+{1\over 2}h_{00}\;,\qquad \Lambda^0_{\;i}={1\over 2}h_{0i}\;,\qquad
\Lambda^i_{\;0}=-{1\over 2}h_{0i}\;,\qquad \Lambda^i_{\;j}=\delta^{i}_{\;j}-{1\over 2}h_{ij}\;.
\end{equation}
A local orthonormal basis $\omega^\alpha_{\;(\beta)}$ is then defined by
\begin{eqnarray}
\nonumber
\omega^\alpha_{\;(0)}&=&\left(1+{1\over 2}h_{00}\;,\;0\;,\;0\;,\;0\right)\;,\qquad\qquad
\omega^\alpha_{\;(1)}=\left(h_{0j}a^j\;,\; a^i-{1\over 2}h_{ij}a^j\right)\;,\\\label{dfgr}\\\nonumber
\omega^\alpha_{\;(2)}&=&\left(h_{0j}b^j\;,\; b^i-{1\over 2}h_{ij}b^j\right)\;,\qquad\qquad
\omega^\alpha_{\;(3)}=\left(h_{0j}k^j\;,\; k^i-{1\over 2}h_{ij}k^j\right)\;.
\end{eqnarray}
By definition, the tetrad frame $e^\alpha_{\;(\beta)}$ is parallel transported along the ray. The propagation equations for these vectors are thus obtained
by applying the operator $D_\lambda$ of the parallel transport
(see (\ref{p1})). Hence,
\begin{equation}
\label{iuk}
{de^\alpha_{\;(\mu)}\over d\lambda}+\Gamma^\alpha_{\beta\gamma}l^\beta e^\gamma_{\;(\mu)}=0\;,
\end{equation}
where $\lambda$ is an affine parameter along the light ray. Using the definition of the Christoffel symbols (\ref{bb1}) and changing over to the variable $\tau$ with $dx^\alpha/d\tau=k^\alpha+O(h)$,
one can recast equations (\ref{iuk}) in the form
\begin{equation}
\label{i9}
{d\over d\tau}\left(e^\alpha_{\;(\mu)}+{1\over 2}h^\alpha_{\;\beta}e^\beta_{\;(\mu)}\right)={1\over 2}\eta^{\alpha\nu}\left(\partial_\nu h_{\gamma\beta}-\partial_\gamma h_{\nu\beta}\right)k^\beta e^\gamma_{\;(\mu)}\;.
\end{equation}
Equation (\ref{i9}) is the main equation for the discussion of the Skrotskii effect.

\subsection{Skrotskii Effect}

We have chosen the null tetrad frame along the ray such that at $t=-\infty$ the tetrad has the property that $e^0_{\;(i)}(-\infty)=0$. It follows that in general $e^0_{\;(i)}=O(h)$.
The propagation equation for the spatial components $e^i_{\;(\mu)}$ is therefore given by
\begin{equation}
\label{i11}
{d\over d\tau}\left(e^i_{\;(\mu)}+{1\over 2}h_{ij}e^j_{\;(\mu)}\right)=
{1\over 2}\left(\partial_i h_{j\beta}-\partial_j h_{i\beta}\right)k^\beta e^j_{\;(\mu)}\;.
\end{equation}
Furthermore, we are interested only in solving equations (\ref{i9}) for the vectors $e^\alpha_{(1)}$ and $e^\alpha_{(2)}$ that are used in the description of the polarization of light; hence
\begin{equation}
\label{i12a}
{d\over d\tau}\left(e^i_{\;(n)}+{1\over 2}h_{ij}e^j_{\;(n)}\right)+\varepsilon_{ijl}e^j_{\;(n)}\Omega^l=0\;,\qquad\qquad (n=1,2)\;,
\end{equation}
where we have defined the quantity $\Omega^i$ as
\begin{equation}
\label{i13}
\Omega^i=-\varepsilon_{ijl}\partial_j\left({1\over 2}h_{l\alpha}k^\alpha\right).
\end{equation}
Therefore, $e^i_{\;(n)}$ can be obtained from the integration of equation (\ref{i12a}). Moreover, $l_\alpha e^\alpha_{\;(n)}=0$ implies that
\begin{equation}
\label{gugu}
e^0_{\;(n)}=k_i e^i_{\;(n)}+h_{0i}e^i_{\;(n)}+h_{ij} k^i e^j_{\;(n)}+\delta_{ij}\dot{\Xi}^i e^j_{\;(n)}\;.
\end{equation}
It is worth noting that as
a consequence of definition $l^\alpha=\omega(1,\dot{x}^i)$ and equation
(\ref{25}), one has ${\bf l}=\omega({\bf k}+\dot{\bm{\Xi}})$. Moreover, from the condition $l_\alpha l^\alpha=0$,
it follows that ${\bf k}\cdot\dot{\bm{\Xi}}=-(1/2)h_{\alpha\beta}k^\alpha k^\beta$.

Let us decompose $\Omega^i$ into components that are parallel and perpendicular to the unit vector $k^i$, i.e.,
\begin{equation}
\label{i14}
\Omega^i=({\bf k}\cdot{\bm{\Omega}})k^i+P^i_{\;j}\Omega^j\;.
\end{equation}
Then, equation (\ref{i12a}) can be expressed as
\begin{equation}
\label{i12b}
{d\over d\tau}\left(e^i_{\;(n)}+{1\over 2}h_{ij}e^j_{\;(n)}\right)+
({\bf k}\cdot{\bm{\Omega}})\varepsilon_{ijl}e^j_{\;(n)}k^l+\varepsilon_{ijl}e^j_{\;(n)}P^l_{\;q}\Omega^q=0\;.\qquad\qquad (n=1,2)
\end{equation}
Integrating this equation from $-\infty$ to $\tau$ taking into account initial conditions (\ref{dqx}) and equalities $\varepsilon_{ijl}a^j k^l=-b^i$ and $\varepsilon_{ijl}b^j k^l=a^i$, we obtain
\begin{eqnarray}
\label{i15}
e^i_{\;(1)}&=&a^i-{1\over 2}h_{ij}a^j+\left(\int^\tau_{-\infty}{\bf k}\cdot{\bm{\Omega}}\;d\sigma\right)b^i-\varepsilon_{ijl}a^jP^l_{\;q}\int^\tau_{-\infty}\Omega^q d\sigma\;,\\\nonumber\\\label{i15a}
e^i_{\;(2)}&=&b^i-{1\over 2}h_{ij}b^j-\left(\int^\tau_{-\infty}{\bf k}\cdot{\bm{\Omega}}\;d\sigma\right)a^i-\varepsilon_{ijl}b^jP^l_{\;q}\int^\tau_{-\infty}\Omega^q d\sigma\;.
\end{eqnarray}

To interpret these results properly, let us note that a rotation in the $(\omega^i_{\;(1)},\omega^i_{\;(2)})$ plane by an angle $\phi$ at time $\tau$ leads to
\begin{eqnarray}
\label{i16}
\omega^i_{\;(1)}&=&(a^i-{1\over 2}h_{ij}a^j)\cos\phi+(b^i-{1\over 2}h_{ij}b^j)\sin\phi\;,\\
\omega^i_{\;(2)}&=&-(a^i-{1\over 2}h_{ij}a^j)\sin\phi+(b^i-{1\over 2}h_{ij}b^j)\cos\phi\;,\label{i17}
\end{eqnarray}
so that if the angle $\phi=O(h)$, then we have
\begin{equation}
\label{i17a}
\omega^i_{\;(1)}=a^i-{1\over 2}h_{ij}a^j+\phi\; b^i\;,\qquad\quad
\omega^i_{\;(2)}=b^i-{1\over 2}h_{ij}b^j-\phi\; a^i\;.
\end{equation}
Comparing these vectors with equations (\ref{i15}) and (\ref{i15a}), we recognize that as vectors ${\bm e}_{(1)}$ and ${\bm e}_{(2)}$ propagate along the light ray they are rotating with an angle
\begin{equation}
\label{i18}
\phi(\tau)=\int^\tau_{-\infty}{\bf k}\cdot{\bm{\Omega}}\;d\sigma
\end{equation}
about ${\bf k}$ in the local $({\bm\omega}_{(1)},{\bm\omega}_{(2)})$ plane. Moreover, ${\bm e}_{(1)}$ and ${\bm e}_{(2)}$ rotate to $O(h)$ toward the direction of light propagation ${\bf k}$ (see the very last terms on the right-hand sides of equations (\ref{i15})and (\ref{i15a})).

We are mostly interested in finding the rotational angle $\phi$ in the plane perpendicular to ${\bf k}$. It is worth noting that the Euclidean dot product
${\bf k}\cdot{\bm{\Omega}}$ can be expressed in terms of partial differentiation
with respect to the impact parameter $\xi^i$ only. This can be done by making use of equation (\ref{pr}) and noting that $\varepsilon_{ijp}k^jk^p\equiv 0$, so that
\begin{equation}
\label{i20}
{\bf k}\cdot{\bm{\Omega}}={1\over 2}k^\alpha k^i\varepsilon_{i\hat{p}\hat{q}}\hat{\partial}_q h_{\alpha\hat{p}}\;,
\end{equation}
where the hat over spatial indices denotes the projection onto the plane
orthogonal to the propagation of light ray, for instance, $A^{\hat{i}}\equiv
P^i_{\;j}A^j$.
Hence, the transport equation for the angle $\phi$ assumes the form
\begin{equation}
\label{x4}
{d\phi\over d\tau}={1\over 2}k^\alpha k^i\varepsilon_{i\hat{p}\hat{q}}\hat{\partial}_q h_{\alpha\hat{p}}\;,
\end{equation}
which is useful for integration. Formula (\ref{x4}) constitutes a significant generalization of a result that was first discussed by Skrotskii (\cite{Sk57} and \cite{Sk68}) and bears his name.

For a stationary "pole-dipole" source of the gravitational field at rest at the origin of coordinates, equations (\ref{chop}), (\ref{2aa}), and (\ref{aqw}) for the metric perturbations imply that
\begin{equation}
\label{i21}
h_{00}={2m\over r}\;,\qquad\qquad
h_{0i}=-{2({\bm{\cal J}}\times {\bf r})^i\over r^3}\;, \qquad\qquad
h_{ij}={2m\over r}\delta_{ij}\;.
\end{equation}
It follows from (\ref{i13}) that in this case
\begin{equation}
\label{i22}
{\bm{\Omega}}={\bf B}_g-{m({\bf k}\times{\bf r})\over r^3}\;,
\end{equation}
where ${\bf B}_g$ is the dipolar gravitomagnetic field associated with the source
\begin{equation}
\label{i23}
{\bf B}_g=\nabla\times\left({{\bm{\cal J}}\times {\bf r}\over r^3}\right)=
{{\cal J}\over r^3}\left[3(\hat{\bf r}\cdot\hat{\bm{\cal J}})\hat{\bf r}-\hat{\bm{\cal J}}\right]\;,
\end{equation}
where $\hat{\bf r}={\bf r}/r$ and $\hat{\bm{\cal J}}={\bm{\cal J}}/{\cal J}$ are unit vectors, and ${\cal J}$ is the magnitude of the angular momentum of the source.
Thus, ${\bf k}\cdot{\bm{\Omega}}={\bf k}\cdot{\bf B}_g$, and, in this way, we recover Skrotskii's original result $d\phi/d\tau={\bf k}\cdot{\bf B}_g$, which is the natural gravitational analog of the Faraday effect in electrodynamics. The Skrotskii effect has a simple physical interpretation in terms of the gravitational Larmor theorem \cite{larmor}.
We note that in the particular case under consideration here, the gravitomagnetic field can be expressed as ${\bf B}_g=-\nabla({\bm{\cal J}}\cdot{\bf r}/r^3)$ for ${\bf r}\not=0$. Thus for radiation propagating in the spacetime exterior to the source, $\phi=-{\bm{\cal J}}\cdot{\bf r}/r^3+constant$. In particular, it follows from this result that the net angle of rotation of the plane of polarization from $-\infty$ to $+\infty$ is zero. This conclusion is confirmed later in equation (\ref{x10}) as well.

In the general case, where bodies generating gravitational field are both moving and rotating, the integration of equation (\ref{x4}) is also straightforward and is accomplished with the
help of a mathematical technique described in Appendix B. The result is
\begin{equation}
\label{x5}
\phi(\tau)=\phi_0+\delta\phi(\tau)-\delta\phi(\tau_0)\;,\qquad\qquad\quad
\delta\phi(\tau)={1\over2}k^\alpha k^i\varepsilon_{i\hat{p}\hat{q}}
\hat{\partial}_q B_{\alpha \hat{p}}(\tau,{\bm{\xi}})\;,
\end{equation}
where $\phi_0$ is a constant angle defining the orientation of the
polarization plane of the electromagnetic wave under discussion at the past null infinity, $\delta\phi(\tau)$ and
$\delta\phi(\tau_0)$ are relativistic rotations of the polarization plane
with respect to its orientation at infinity at the moments of observation,
$\tau$, and emission of light, $\tau_0$,
respectively, and the function $B_{\alpha\beta}$ is defined in (\ref{f1}).

We note that according to the definition (\ref{f1}) the tensor function
$B_{\alpha\beta}$ consists of two parts the first of which
$B^{\alpha\beta}_M$ relates to the action of the mass monopole field of
the particles and the second one $B^{\alpha\beta}_S$ describes the action
of spin dipole fields on the rotation of the polarization plane.
Therefore, the angle of rotation can be represented as an algebraic sum of
two components
\begin{eqnarray}
\label{x6}
\delta\phi&=&\delta\phi_M+\delta\phi_S\;,
 \end{eqnarray}
where the monopole part can be calculated using equation (\ref{f5b}),
\begin{eqnarray}
\label{x6a}
\delta\phi_M&=&{1\over2}k_\alpha k_i\varepsilon^{i\hat{p}\hat{q}}
\hat{\partial}_q B^{\alpha \hat{p}}_M=
2m\;{1-{\bf k}\cdot{\bf v}\over\sqrt{1-v^2}}
\frac{{\bf k}\cdot({\bf v}\times{\bm{\xi}})}{(r-{\bf k}\cdot{\bf r})(r-{\bf
v}\cdot{\bf r})}\;.
\end{eqnarray}
The spin part is given by
\begin{eqnarray}
\label{x6b}
\delta\phi_S&=&{1\over2}k_\alpha k_i\varepsilon^{i\hat{p}\hat{q}}
\hat{\partial}_q B^{\alpha \hat{p}}_S\;,
\end{eqnarray}
and has a rather complicated form if written explicitly by
making use of partial derivative of
function $B^{\alpha\beta}_S$ given in (\ref{ff5}).
In the gravitational lens approximation when light goes from $-\infty$ to $+\infty$, equations (\ref{x6a}) and (\ref{x6b})
simplify and assume the form
\begin{equation}
\label{x9}
\delta\phi_M(+\infty)
=-{4Gm\over c^3d^2}{\bf k}\cdot({\bm{\xi}}\times{\bf v})\;,
\end{equation}
\begin{equation}
\label{x10}
\delta\phi_S(+\infty)
=\frac{4G}{c^3 d^2}\left[({\bf k}\cdot{\bm{\cal J}})+
\frac{({\bf k}\times{\bm{\xi}})\cdot({\bm{\xi}}\times{\bm{\cal J}})}
{d^2}\right]\equiv 0\;.
\end{equation}

Equations (\ref{x9}) and (\ref{x10}) make it evident that in the case of
gravitational lensing the integrated effect discussed by Skrotskii \cite{Sk57}
is only due to the translational
motion of the lens. There is no contribution to the effect caused by
spin of the body and proportional to $1/d^2$, where $d$ is the impact
parameter of light ray with respect to the body. This rather remarkable
fact was first noted by
Kobzarev \& Selivanov \cite{Kob} who criticized final conclusions of the
paper by Skrotskii \cite{Sk57} and others (see \cite{er2} and \cite{er3}). We
would like to stress, however, that all of the results of the paper by Skrotskii
\cite{Sk57} are correct, since the final formulas of his paper relate to the
situation where light is emitted from or near the surface of the rotating body.
Such a case obviously is not reduced to that of gravitational lens. Thus,
the angle of rotation of the polarization plane does not equal to zero as
correctly shown by Skrotskii.

The absence of the Skrotskii effect for radiation propagating from $-\infty$ to $+\infty$ in the case of a stationary rotating body 
reveals that the coupling of the polarization vector of the photon with the
spin of the rotating body can not be
amplified by the presence of a gravitational lens.

\section{The Skrotskii Effect by Quadrupolar Gravitational Waves from Localized Sources}
\subsection{Quadrupolar Gravitational Wave Formalism}

We shall consider in this section the Skrotskii effect associated with the
emission of gravitational waves from a localized astronomical system like
a binary star, supernova explosion, etc. For simplicity we shall restrict
our considerations to the quadrupole approximation only. The direct way to
tackle the problem would be to use the Taylor expansion of the formula for the
Skrotskii effect given in the
previous section. However, it is instructive
to make use of an approach outlined in \cite{KSGE} that is based on the
multipole expansion of radiative gravitational field of a localized
astronomical source \cite{BKS}.

To this end, let us consider the propagation of light ray taking place always
outside the source with its center of mass at rest that emits gravitational waves.
The metric perturbation $h_{\mu\nu}$ can be
split into a canonical part $h^{can.}_{\mu\nu}$,
which contains symmetric trace-free (STF) tensors only (for more details on
STF tensors see \cite{BDam}), and a gauge part, i.e.
$h_{\mu\nu} = h^{can.}_{\mu\nu} + \partial_{\mu}w_{\nu} +
\partial_{\nu}w_{\mu}$. In the case of mass-monopole, spin-dipole,
and mass-quadrupole source moments and in the harmonic gauge the canonical
part of the metric perturbation is given by 
\begin{eqnarray}
\label{b21}
h_{00}^{can.}&=&\frac{2{\cal M}}{r}+\partial_{pq}
\left[\frac{{\cal I}_{pq}(t-r)}{r}\right]\;,\\\nonumber\\\label{b22}
h_{0i}^{can.}&=&-\frac{2\varepsilon_{ipq}{\cal J}_p x_q}{r^3}+
2\partial_j\left[\frac{\dot{\cal I}_{ij}(t-r)}{r}\right]\;,\\\nonumber\\
\label{b23}
h_{ij}^{can.}&=&\delta_{ij}h_{00}^{can.}+\frac{2}{r}\ddot{\cal I}_{ij}(t-r)\;.
\end{eqnarray}
Here the mass ${\cal M}$ and spin ${\cal J}^i$ of the source are constants,
whereas its quadrupole moment ${\cal I}_{ij}$ is a function
of the retarded time $t-r$. Dependence
of the mass and spin on time would be caused by the process of emission of energy
and angular momentum that are then carried away from the source by gravitational waves.
If necessary, the time dependence of mass and spin can be treated in the
framework of the same calculational scheme as applied in the present
paper. Moreover, it is assumed in (\ref{b21}) -- (\ref{b23}) that the center of mass of the source of
gravitational waves is at the origin of the coordinate system and does not
move. Hence, the radial coordinate $r$ is the distance from the center of mass
to the field point in space.

The explicit expressions for the gauge functions
$w^{\mu}$ relating $h^{can.}_{\mu\nu}$ with $h_{\mu\nu}$ are chosen in the
form \cite{KS}
\begin{eqnarray}\label{ttt}
w^0&=& \frac{1}{2}\nabla_k \nabla_l \left[
\frac{^{(-1)}{\cal{I}}_{kl}(t-r)}{r}\right]\;,
\\\nonumber\\\label{kkk}
w^i&=&\frac{1}{2}  \nabla_i \nabla_k \nabla_l\left[
\frac{^{(-2)}{\cal{I}}_{kl}(t-r)}{r}\right] -
2\nabla_k \left[\frac{{\cal{I}}_{ki}(t-r)}{r}\right]\;,
\end{eqnarray}
where we have introduced the definitions of time integrals of the quadrupole moment
\begin{equation}\label{integr}
^{(-1)}{\cal{I}}_{ij}(t)\equiv\int_{-\infty}^t d\upsilon\;{\cal{I}}_{ij}(\upsilon)\;,
\quad\quad
^{(-2)}{\cal{I}}_{ij}(t)\equiv\int_{-\infty}^t
d\upsilon\;^{(-1)}{\cal{I}}_{ij}(\upsilon)\;.
\end{equation}

Gauge functions (\ref{ttt}) and (\ref{kkk}) make space-time coordinates
satisfy both harmonic and ADM gauge conditions
simultaneously \cite{KSGE}. The harmonic-ADM coordinates are especially
useful for integrating equations of light propagation and the description of
the motions of free falling source of light and observer. It turns out that
the
source of light and the observer do not experience the influence of gravitational waves
in the ADM coordinates and move only under the action of the stationary part of the
gravitational field created by the mass and spin of the source of the
gravitational waves (for more details concerning the construction of the
harmonic-ADM coordinates see \cite{KSGE}).

The rotation of the polarization plane is described by equation (\ref{x4}),
where the rotation frequency $\Omega$ is defined in
(\ref{i13}). Using expressions (\ref{b21}) - (\ref{b23})
together with the gauge freedom, the angular velocity is given by
\begin{eqnarray}
\label{sd1}
\Omega&=&\hat\partial_{\tau}\left(-{{\cal J}_i x^i\over r^3}\right)+
k^i\varepsilon_{i\hat{p}\hat{q}}\hat\partial_{qj}
\left[\frac{\dot{\cal I}_{jp}(t-r)}{r}\right]+
k^i\varepsilon_{i\hat{p}\hat{q}}\hat\partial_{q\tau}\left[
\frac{k^j\dot{\cal I}_{jp}(t-r)}{r}\right]-
\hat\partial_\tau\left[{1\over2}\;{\bf
k}\cdot\left({\bm{\nabla}}\times{\bf w}\right)\right]\;,
\end{eqnarray}
where ${\bf w}\equiv (w^i)$ is given by (\ref{kkk}), $r=\sqrt{\tau^2+d^2}$,
and $d=|{\bm{\xi}}|$ is the impact parameter of the light ray.

Integration of equation (\ref{x4}) with respect to time results in
\begin{eqnarray}
\label{sd2}
\delta\phi(\tau)&=&
-\frac{{\cal J}_i x^i}{r^3}-
{1\over2}\;{\bf k}\cdot\left({\bm{\nabla}}\times{\bf B}\right)\;,
\end{eqnarray}
where ${\bf B}\equiv (B^i)$ with 
\begin{eqnarray}
\label{sd3}
B^i&=&\frac{2\xi^j\dot{\cal I}_{ij}(t-r)}{ry}+
\frac{2k^j\dot{\cal I}_{ij}(t-r)}{r}-2\nabla_k \left[\frac{{\cal{I}}_{ki}(t-r)}{r}\right]
\end{eqnarray}
and $y=\tau-r$. Making use of formula (\ref{kkk}) for $w^i$ and taking
partial derivatives brings equation (\ref{sd2}) to the following
more explicit form
\begin{eqnarray}
\label{sd4}
\delta\phi(\tau)&=&-\frac{{\cal J}_i x^i}{r^3}+\frac{({\bf k}\times{\bm{\xi}})^i \xi^j}{ry}\left[
-\frac{\dot{\cal I}_{ij}}{ry}+\frac{\dot{\cal I}_{ij}}{r^2}+
\frac{\ddot{\cal I}_{ij}}{r}\right]
+\frac{({\bf k}\times{\bm{\xi}})^i x^j}{r^3}\left[
\ddot{\cal I}_{ij}+\frac{3\dot{\cal I}_{ij}}{r}+
\frac{3{\cal I}_{ij}}{r^2}\right]
+\frac{({\bf k}\times{\bm{\xi}})^i k^j}{r^2}\left[
\ddot{\cal I}_{ij}+\frac{\dot{\cal I}_{ij}}{r}\right],
\end{eqnarray}
where the quadrupole moments ${\cal I}_{ij}$ are all evaluated at the retarded
time $s=t-r$.
It is straightforward to obtain two limiting cases of this formula related
to the cases of a gravitational lens and a plane gravitational wave.

\subsection{Gravitational Lens Approximation}

In the
event of gravitational lensing one has \cite{KSGE}
\begin{equation}
\label{sd5}
y\equiv\tau-r=\sqrt{r^2-d^2}-r=-{d^2\over2r}+...\;,
\end{equation}
so that
\begin{equation}
\label{sd5a}
ry=-{d^2\over2}\;,\qquad\qquad\qquad
t-r=t^\ast-{d^2\over2r}+...\;,
\end{equation}
where $t^\ast$ is the instant of the closest approach of light ray to the
center of mass of the gravitational lens. After making use of the
approximations shown above,
equation (\ref{sd4}) simplifies and assumes the form
\begin{eqnarray}
\label{sd6}
\delta\phi(t^\ast)&=&-\frac{4\dot{\cal I}_{ij}(t^\ast)
({\bf k}\times{\bm{\xi}})^i \xi^j}{d^4}\;.
\end{eqnarray}
One can see that the first
nonvanishing contribution to the Skrotskii effect comes from the time
derivative of a mass quadrupole moment and does not depend on the spin of the source of
gravitational waves. It agrees with the conclusions of section 7.
Formula (\ref{sd6}) also shows that if the source of gravitational waves
is periodic like a binary star system the polarization plane of the
electromagnetic wave will
experience periodic changes of its orientation with a characteristic
frequency that is twice that of the source \cite{Kob}.

Equation (\ref{sd6}) can be derived from the equation (\ref{x9}) as well.
Indeed, let us assume that the gravitational lens is comprised of N point
particles forming a self-gravitating body. For N point particles,
equation (\ref{x9}) implies
\begin{eqnarray}
\label{sd6a}
\delta\phi_M(+\infty)
&=&-4\sum_{a=1}^N {Gm_a\over c^3|{\bm{\xi}}-{\bm{\xi}}_a|^2}
{\bf k}\cdot[({\bm{\xi}}-{\bm{\xi}}_a)\times{\bf v}_a]\;,
\end{eqnarray}
where the point particles are enumerated by index `a' running from 1 to N,
$m_a$ is mass of the $a$th particle, ${\bf x}_a$ and ${\bf v}_a$ are
coordinates and velocity of the $a$th particle, $\xi^i_a=P^i_{\;j}x^j_a$,
and $\xi^i$ is the impact parameter of the light ray with respect to the
origin of coordinate system that we assume coincides with the center of mass
of the lens, ${\cal I}^i$, defined by the equation
\begin{eqnarray}
\label{sd6b}
{\cal I}^i(t^\ast)&=&\sum_{a=1}^N m_a x^i_a(t^\ast)=0\;.
\end{eqnarray}
As a consequence of (\ref{sd6b}), we conclude that all of the time derivatives of
${\cal I}^i$ vanish identically. We also define the spin of the lens and its
tensor of inertia as follows
\begin{equation}
\label{sd6c}
{\cal J}^i=\sum_{a=1}^N m_a ({\bf x}_a\times{\bf
v}_a)^i\;,\qquad\qquad
I^{ij}(t^\ast)=\sum_{a=1}^N m_a x^i_a(t^\ast)x^j_a(t^\ast)\;.
\end{equation}
The spin of the lens is conserved and does not depend on time, while the tensor
of inertia is, in general, a function of time.

Let us expand the right-hand side of equation (\ref{sd6a}) in a power series
with respect to the parameter $|{\bm{\xi}}_a|/|{\bm{\xi}}|$ which is
presumed to be small. We find that
\begin{eqnarray}
\label{sd6d}
\delta\phi_M(+\infty)
&=&-{4G\over c^3d^2}\left[{\bf k}\cdot\left({\bm{\xi}}\times\sum_{a=1}^N m_a {\bf
v}_a\right)-
{\bf k}\cdot\sum_{a=1}^N m_a \left({\bm{\xi}}_a\times{\bf
v}_a\right)\right]-
{8({\bf k}\times{\bm{\xi}})^i\over d^4} \sum_{a=1}^N m_a
v^i_a({\bm{\xi}}_a\cdot{\bm{\xi}})+O\left(v_a{\xi_a\over d}\right)^2\;,
\end{eqnarray}
where $d=|{\bm{\xi}}|$. Taking into account definitions of the center of
mass, spin, tensor of inertia of the lens, as well as the identity
\begin{eqnarray}
\label{sd6e}
2[({\bf k}\times{\bm{\xi}})\cdot{\bf v}_a]({\bm{\xi}}_a\cdot{\bm{\xi}})
&=&({\bf k}\times{\bm{\xi}})^i\left[\dot{\cal
I}_{ij}\xi^j-({\bm{\xi}}\times{\bm{\cal J}})^i\right]
\end{eqnarray}
together with formula (\ref{x9}) we arrive at the result shown in (\ref{sd6}). Comparing expression
(\ref{sd6}) with the corresponding result derived by Kobzarev \& Selivanov
(\cite{Kob}, formulas (11) and (12)), we conclude that the result given by these authors \cite{Kob}
is misleading.

\subsection{Plane Gravitational Wave Approximation}

In the limit of a plane gravitational wave, we assume that
the distance $D$ between the observer and source
of light is much smaller than $r$ and $r_0$, their respective distances from
the deflector. The following exact equalities hold:
\begin{equation}
\label{q2}
d=r\sin\vartheta\;,\qquad\quad
y=\tau-r=r(\cos\vartheta-1)\;,
\end{equation}
where $\vartheta$ is the angle between the directions from the observer to the source of
light and the source of gravitational waves.
Furthermore,
we note that the vector ${\bm\xi}$, corresponding to impact parameter $d$,
can be represented as
\begin{eqnarray}\label{vecxi}
\xi^i&=&r\left(N^i-k^i\cos\vartheta\right)\;,
\end{eqnarray}
where $N^i=x^i/r$ and $|{\bf N}|=1$. Making asymptotic expansions up to leading
terms of order $1/r$ and $1/r_0$, and neglecting all residual terms of
order $1/r^2$ and $1/r^2_0$, would lead to the following result
\begin{eqnarray}
\label{sd7}
\delta\phi(\tau)&=&\frac{({\bf k}\times{\bf N})^i(N^j\cos\vartheta-k^j)\ddot
{\cal I}_{ij}(t-r)}{r(\cos\vartheta-1)}=
\frac{({\bf k}\times{\bf N})^i k^j\ddot
{\cal I}^{TT}_{ij}(t-r)}{r(1-\cos\vartheta)}\;,
\end{eqnarray}
where the definition of ``transverse-traceless" tensor \cite{MTW}
with respect to the direction ${\bf N}$ is taken into account
\begin{eqnarray}\label{TT}
{\cal I}_{ij}^{TT}&=&{\cal I}_{ij}+
\frac{1}{2}\left(\delta_{ij}+N_i N_j\right)N_p N_q\; {\cal
I}_{pq}-\left(\delta_{ip}N_j N_q+\delta_{jp}N_i N_q\right)\;{\cal I}_{pq}\;,
\end{eqnarray}
and the projection is onto the plane orthogonal to the unit vector ${\bf N}$.
In terms of the transverse-traceless metric perturbation $h^{TT}_{ij}=2\ddot{\cal
I}^{TT}_{ij}(t-r)/r$, the angle of rotation of the polarization plane of
electromagnetic wave emitted at past null infinity is given by
\begin{eqnarray}
\label{sd8}
\delta\phi(\tau)&=&{1\over2}
\frac{({\bf k}\times{\bf N})^i k^j}{1-\cos\vartheta}\;h^{TT}_{ij}(t-r)\;.
\end{eqnarray}
In the case of electromagnetic wave emitted at instant $t_0$ and at the distance
$r_0$ from the source of gravitational waves and received at instant $t$
at the distance $r$, the overall angle of rotation of the polarization plane
is defined as a difference
\begin{eqnarray}
\label{sd8a}
\delta\phi(\tau)-\delta\phi(\tau_0)&=&{1\over2}\left[\frac{({\bf k}\times{\bf N})^i k^j}{1-\cos\vartheta}\;h^{TT}_{ij}(t-r)-\frac{({\bf k}\times{\bf N}_0)^i k^j}{1-\cos\vartheta_0}\;h^{TT}_{ij}(t_0-r_0)  \right]\;,
\end{eqnarray}
where $N_0^i=x_0^i/r_0$ and $\vartheta_0$ is the angle between the vectors ${\bf k}$
and ${\bf N}_0$.

It is worthwhile to compare the results obtained in the present section
to the standard approach used for calculating effects caused
by plane gravitational waves.
Ordinarily in such an approach a plane gravitational wave
is assumed to be monochromatic with the property that at infinity the space-time
remains asymptotically flat. Actually it means that one deals with a localized
packet of the waves with boundaries asymptotically approaching to plus and
minus null infinity
uniformly. The metric tensor of such a plane gravitational wave in the
transverse-traceless (TT) gauge can be expressed as \cite{MTW}
\begin{equation}
\label{pw1}
h_{00}=h_{0i}=0\;\qquad\quad h^{TT}_{ij}={\rm Re}\{\hat{a}_{ij}\exp(i\hat{p}_\alpha
x^\alpha)\}\;,
\end{equation}
where the spatial tensor $\hat{a}_{ij}$ is symmetric and traceless ($\hat{a}_{kk}=0$),
$\hat{p}_\alpha=\omega_{gr}(-1,\hat{\bf p})$ is the propagation 4-vector of the
gravitational wave with constant frequency $\omega_{gr}$ and constant unit
spatial propagation vector $\hat{\bf p}$, and $\hat{a}_{ij}\hat{p}^j=0$ \cite{aij}. One can easily see that
such a localized plane gravitational wave does not interact at all with an
electromagnetic ray propagating from minus to plus null infinity.

Indeed, the integration of the light ray equation (\ref{14a}) leads to integrals involving the exponential function entering (\ref{pw1}) along the unperturbed
light ray path of the form
\begin{equation}
\label{pw2}
\lim_{T\to \infty}\int^{+T}_{-T}e^{i\hat{p}_\alpha
k^\alpha\tau}d\tau=2\pi\delta(\hat{p}_\alpha k^\alpha)=2\pi
\omega^{-1}_{gr}\delta({\bf k}\cdot\hat{\bf p}-1)\;,
\end{equation}
where $\delta(x)$ is the Dirac delta function.
Now, if the direction of propagation of electromagnetic ray does not
coincide with that of the gravitational wave (${\bf k}\neq\hat{\bf p}$) the
argument of the delta
function in (\ref{pw2}) is not zero, i.e. ${\bf k}\cdot\hat{\bf p}<1$, and the
result of integration along
the light ray trajectory vanishes. On the other hand, if ${\bf k}=\hat{\bf p}$ there is again no effect since the right side of equation (\ref{14a}) vanishes due to the transversality of $h^{TT}_{ij}$
revealing that in the case under consideration $h^{TT}_{ij}k^j=
h^{TT}_{ij}\hat{p}^j=0$. This result has been proved by Damour and
Esposito-Far\`{e}se \cite{DamE} who studied the deflection of light and
the integrated 
time delay caused by the time-dependent gravitational field generated by a
localized material source lying close to the line of sight \cite{Zeld} --\cite{brag}.

The absence of interaction of plane monochromatic gravitational waves with
electromagnetic rays propagating from minus to plus null infinity makes it
evident that all relativistic effects taking place in the gravitational
lens approximation have little to do with gravitational waves --- i.e., only the near-
zone gravitational field of the lens contributes to the overall effects.
The other conclusion is that the plane gravitational wave disturbs the 
propagation of electromagnetic signals if and only if the signals go to
finite distances \cite{Mash2} -\cite{MashG}. However, another limitation
has to be kept in mind, namely, as it follows for instance from (\ref{sd8a})
the standard plane gravitational wave approximation (\ref{pw1}) is applicable
only for gravitational waves with dominant wavelengths $\lambda_{gr}\ll\min(r,r_0)$.
This remark is especially important for having a self-consistent analysis
of such intricate
problems as the detection of the solar g-modes by interferometric
gravitational wave detectors \cite{gwd}, the theoretical prediction of
low-frequency pulsar timing noise (see, e.g., \cite{ptn1} --\cite{ptn4},
and references therein), and the anisotropy of
cosmic microwave background radiation \cite{cmb} caused by primordial
gravitational waves.

\appendix
\section{Geometric Optics Approximation from the Maxwell Equations}

Here we give a brief derivation of the equations of geometric optics
approximation, discussed in section 4, from the Maxwell equations for
electromagnetic waves propagating in empty space-time. The
source-free Maxwell equations are given by \cite{landau}
\begin{equation}
\label{m1a}
\nabla_\alpha F_{\beta\gamma}+\nabla_\beta F_{\gamma\alpha}+\nabla_\gamma
F_{\alpha\beta}=0\;,
\end{equation}
\begin{equation}
\label{m1b}
\nabla_\beta F^{\alpha\beta}=0\;,
\end{equation}
where $\nabla_\alpha$ denotes covariant differentiation \cite{note3}. Taking a
covariant derivative from equation (\ref{m1a}) and using
equation (\ref{m1b}), we obtain the covariant wave equation for the
electromagnetic field tensor
\begin{equation}
\label{m2}
{\dAl}_g F_{\alpha\beta}+R_{\alpha\beta\gamma\delta}F^{\gamma\delta}-R_{\alpha\gamma}F^\gamma_{\;\;\beta}+R_{\beta\gamma}R^\gamma_{\;\;\alpha}=0\;,
\end{equation}
where ${\dAl}_g\equiv\nabla^\alpha\nabla_\alpha$, $R_{\alpha\beta\gamma\delta}$ is the Riemann curvature tensor, and $R_{\alpha\beta}=R^\gamma_{\;\;\alpha\gamma\beta}$ is the Ricci tensor.

Let us now assume that the electromagnetic tensor $F_{\alpha\beta}$ has a
form shown in equation (\ref{ff1}). We introduce a dimensionless perturbation
parameter $\varepsilon$ and assume an expansion of the field of the form
\begin{equation}
\label{m3}
F_{\alpha\beta}={\rm Re}\left[\left(a_{\alpha\beta}+ \varepsilon
b_{\alpha\beta}+\varepsilon^2
c_{\alpha\beta}+...\right)\exp\left({i\varphi\over\varepsilon}\right)\right]\;;
\end{equation}
see \cite{mash87} for a critical examination of this procedure and its underlying physical assumptions.
Substituting the expansion (\ref{m3}) into equation (\ref{m1a}), taking into account
the definition $l_\alpha=\partial\varphi/\partial x^\alpha$, and arranging
terms with similar powers of $\varepsilon$ lead to the chain of equations
\begin{eqnarray}
\label{m4a}
l_\alpha\; a_{\beta\gamma}+l_\beta\; a_{\gamma\alpha}+l_\gamma\;
a_{\alpha\beta}&=&0\;,\\
\label{m4b}
i\left(\nabla_\alpha\; a_{\beta\gamma}+\nabla_\beta\; a_{\gamma\alpha}+\nabla_\gamma\;
a_{\alpha\beta}\right)&=&
l_\alpha\; b_{\beta\gamma}+l_\beta\; b_{\gamma\alpha}+l_\gamma\;
b_{\alpha\beta}\;,
\end{eqnarray}
etc., where we have assumed the effects of curvature are negligibly small . Similarly, equation (\ref{m1b}) gives a chain of equations
\begin{eqnarray}
\label{m5a}
l_\beta\; a^{\alpha\beta}&=&0\;,\\
\label{m5b}
\nabla_\beta\; a^{\alpha\beta}+il_\beta\; b^{\alpha\beta}&=&0\;,
\end{eqnarray}
and so on.

Equation (\ref{m5a}) implies that the electromagnetic field tensor is
orthogonal in the four-dimensional sense to vector $l_\alpha$.
Contracting equation (\ref{m4a}) with $l_\alpha$ and accounting for
(\ref{m5a}), we find that $l_\alpha$ is null, $l_\alpha l^\alpha=0$.
Taking the covariant derivative of this equality and using the fact that $\nabla_{[\beta}\;l_{\alpha]}=0$ since $l_\alpha=\nabla_\alpha\varphi$, one can show that the vector
$l_\alpha$ obeys the null geodesic equation (\ref{p0}). Finally, equation (\ref{m2}) can be used to show that
\begin{equation}
\label{m6}
\nabla_\gamma(a_{\alpha\beta}\;l^\gamma)+l^\gamma\nabla_\gamma\; a_{\alpha\beta}=0\;,
\end{equation}
which immediately leads to equation (\ref{p1}) of the propagation law
for the electromagnetic field tensor. In this way, one can prove the validity of the equations of
the geometric optics approximation displayed in section 4.

\section {The Approximate Expressions for the Christoffel Symbols}

Making use of the general definition of the Christoffel symbols \cite{MTW}
\begin{equation}
\label{bb1}
\Gamma^\alpha_{\beta\gamma}={1\over2}g^{\alpha\delta}\left(
\partial_\gamma\; g_{\delta\beta}+\partial_\beta\; g_{\delta\gamma}-
\partial_\delta\; g_{\beta\gamma}\right)\;,\quad\quad\quad\quad
\partial_\alpha\equiv \partial/\partial x^\alpha\;,
\end{equation}
and applying the expansion of the metric tensor (\ref{2a}) result in the
approximate  post-Minkowskian expressions
\begin{eqnarray}
\label{bb2}
\Gamma^0_{00}&=&-{1\over2}\partial_t h_{00}(t,{\bf x})\;,\qquad\qquad
\partial_t\equiv {\partial\over \partial t}\;,\\
\label{bb3}
\Gamma^0_{0i}&=&-{1\over2}\partial_i h_{00}(t,{\bf x})\;,\qquad\qquad
\partial_i\equiv {\partial\over \partial x^i}\;,\\
\label{bb4}
\Gamma^0_{ij}&=&-{1\over2}\left[\partial_i h_{0j}(t,{\bf x})+
\partial_j h_{0i}(t,{\bf x})-\partial_t h_{ij}(t,{\bf x})\right]\;,\\
\label{bb5}
\Gamma^i_{00}&=&\partial_t h_{0i}(t,{\bf x})-{1\over2}
\partial_i h_{00}(t,{\bf x})\;,\\
\label{bb6}
\Gamma^i_{0j}&=&{1\over2}\left[\partial_j h_{0i}(t,{\bf x})-
\partial_i h_{0j}(t,{\bf x})+\partial_t h_{ij}(t,{\bf x})\right]\;,\\
\label{bb7}
\Gamma^i_{jp}&=&{1\over2}\left[\partial_j h_{ip}(t,{\bf x})+
\partial_p h_{ij}(t,{\bf x})-\partial_i h_{jp}(t,{\bf x})\right]\;.
\end{eqnarray}
These expressions are used for the calculation of the right-hand side of
the null geodesic equation (\ref{ff2}).

\section{Calculation of Integrals along the Light-Ray Trajectory}

In order to calculate the integrals (\ref{f1}) and (\ref{f2}), it
is useful to change in the integrands the time argument, $\sigma$,
to the new one, $\zeta$, defined by the light-cone equation
(\ref{fgh2}), which after substituting the unperturbed light
trajectory (\ref{11a}) for ${\bf x}$ can be expressed as follows
\begin{equation}
\label{17} \sigma+t^{\ast}=\zeta+|{\bm {\xi}}+{\bf k}\sigma-{\bf
z}(\zeta)|\;,
\end{equation}
where $t^{\ast}$, ${\bm {\xi}}$, and ${\bf k}$ are considered as
parameters such that the orthogonality condition $\xi_i k^i=0$ 
holds. The differentiation of equation (\ref{17}) yields the
partial derivatives of the retarded time with respect to the
parameters
\begin{equation}
\label{diff}
\frac{\partial \zeta}{\partial t^\ast}=\frac{r}{r-{\bf v}\cdot{\bf
r}}\;,\quad\quad\quad
P_{ij}\frac{\partial \zeta}{\partial \xi^j}=-\frac{P_{ij}r^j}{r-{\bf v}\cdot{\bf
r}}\;,\quad\quad\quad
\frac{\partial \zeta}{\partial k^i}=-\frac{\sigma r^i}{r-{\bf v}\cdot{\bf
r}}\;,
\end{equation}
and the relationship between the time differentials along the world-line of the photon
\begin{equation}
\label{newvar}
d\sigma=d\zeta\;\frac{r-{\bf v}\cdot{\bf r}}{r-{\bf k}\cdot{\bf
r}}\;.
\end{equation}
It is worth noting that in the formula (\ref{diff}) one can write
$P_{ij}r^j=({\bf k}\times({\bf r}\times{\bf k}))^i\equiv
r^i-k^i({\bf k}\cdot{\bf r})$. Moreover, $\xi_i=P_{ij}\xi^j$ since
$P^{ij}=\delta^{ij}-k^i k^j$ and $k_i\xi^i=0$, i.e. $\xi^i$ has
only two independent components. 

Two principal differential identities applied to any smooth
function $F(t,{\bf x})$ are used in calculations throughout the
paper, namely, as proved in \cite{KSGE}
\begin{eqnarray}
\label{pr}
\left[\frac{\partial F(t,{\bf x})}{\partial x^i}+
k_i\;\frac{\partial F(t,{\bf x})}{\partial t}\right]_{{\bf x}={\bf
k}(t-t_0)+{\bf x}_0}&=&P_{ij}\frac{\partial F(\tau+t^\ast,{\bf k}\tau+{\bm{\xi}})}
{\partial \xi^j}+
k_i\;\frac{\partial F(\tau+t^\ast,{\bf k}\tau+{\bm{\xi}})}{\partial \tau}\;,
\end{eqnarray}
and
\begin{eqnarray}
\label{note}
\left[\frac{\partial F(t, {\bf x})}{\partial t}
\right]_{t=\sigma+t^\ast;\;{\bf x}={\bf k}\sigma+{\bm{\xi}}}&=&
\frac{\partial F(\sigma+t^\ast, {\bf k}\sigma+{\bm{\xi}})}
{\partial t^\ast}\;.
\end{eqnarray}
Relationship (\ref{pr}) allows one to change the order of
operations of partial differentiation and substitution for the
unperturbed light ray trajectory, while (\ref{note}) shows how to
change differentiation from the time $t$ to the parameter
$t^{\ast}$ making use of the reparametrization of the light-ray
trajectory. Parameters $\xi^i$ and $t^{\ast}$ do not depend on
time and, for this reason, derivatives with respect to them can be
taken out of the integrals along the light-ray trajectory. Hence,
applying (\ref{pr}) and (\ref{note}) to the integral (\ref{f1})
and accounting for (\ref{newvar}) we obtain
\begin{eqnarray}
\label{f3}
B^{\alpha\beta}_S(\tau,{\bm{\xi}})&=&4\frac{\partial}{\partial t^\ast}
\int^{s(\tau,t^\ast)}_{-\infty}\frac{k_\gamma S^{\gamma(\alpha}u^{\beta)}}{r-{\bf
k}\cdot{\bf r}}d\zeta-4\frac{\partial}{\partial\xi^i}\int^{s}_{-\infty}
\frac{S^{i(\alpha}u^{\beta)}}{r-{\bf k}\cdot{\bf r}}d\zeta
-4\frac{k_i S^{i(\alpha}u^{\beta)}}{r-{\bf v}\cdot{\bf r}}\;.
\end{eqnarray}
where the expression
\begin{eqnarray}
\label{f4}
r-{\bf k}\cdot{\bf r}&=&t^\ast+{\bf k}\cdot{\bf z}(\zeta)-\zeta\;,
\end{eqnarray}
is a function of the retarded argument $\zeta$ and the (constant)
parameter $t^\ast$ only. Differentiation with respect to the
parameters $t^\ast$ and $\xi^i$ in equation (\ref{f3}) results in
\begin{eqnarray}
\label{f5}
B^{\alpha\beta}_S(\tau,{\bm{\xi}})&=&
\frac{4r_\gamma S^{\gamma(\alpha}u^{\beta)}}{(r-{\bf v}\cdot{\bf r})
(r-{\bf k}\cdot{\bf r})}-
4\int^{s}_{-\infty}\frac{k_\gamma S^{\gamma(\alpha}u^{\beta)}}
{(r-{\bf k}\cdot{\bf r})^2}d\zeta\;,
\end{eqnarray}
where $r^\alpha=(r,r^i)$ and $r_\alpha=(-r,r^i)$. A useful formula
is obtained after differentiating $B^{\alpha\beta}_S$ with respect
to the impact parameter. Specifically, we have
\begin{eqnarray}
\label{ff5}
\hat{\partial}_i B^{\alpha\beta}_S&=&
\frac{4P_{ij}S^{j(\alpha}u^{\beta)}}{(r-{\bf v}\cdot{\bf r})
(r-{\bf k}\cdot{\bf r})}+
\frac{4r_\gamma S^{\gamma(\alpha}u^{\beta)}}{r-{\bf k}\cdot{\bf r}}
\frac{P_{ij}v^j}{(r-{\bf v}\cdot{\bf r})^2}\nonumber\\\nonumber\\
&+&4P_{ij}r^j\;\left[
\frac{k_\gamma S^{\gamma(\alpha}u^{\beta)}}{(r-{\bf v}\cdot{\bf r})
(r-{\bf k}\cdot{\bf r})^2}-
\frac{(1-v^2)r_\gamma S^{\gamma(\alpha}u^{\beta)}}{(r-{\bf k}\cdot{\bf
r})(r-{\bf v}\cdot{\bf r})^3}-
\frac{(1-{\bf k}\cdot{\bf v})r_\gamma S^{\gamma(\alpha}u^{\beta)}}
{(r-{\bf v}\cdot{\bf r})^2(r-{\bf k}\cdot{\bf r})^2}\right]\;.
\end{eqnarray}
For comparison, in the case of pure monopole particles we have
\cite{KS}
\begin{eqnarray}
\label{f5b}
\hat{\partial}_i B^{\alpha\beta}_M(\tau)&=&-4m(1-v^2)^{1/2}
\frac{u^\alpha u^\beta+{1\over2}\eta^{\alpha\beta}}{r-{\bf v}\cdot{\bf r}}
\frac{P_{ij}r^j}{r-{\bf k}\cdot{\bf r}}\;.
\end{eqnarray}
As for the integral $D^{\alpha\beta}_S$, one can see from
(\ref{27}) and (\ref{28}) that in the calculation of the light-ray
trajectory we do not need $D^{\alpha\beta}$ directly but only its
partial derivative with respect to the parameter ${\xi}^i$. Thus,
using the definition of $D^{\alpha\beta}_S$ and the expression for
$B^{\alpha\beta}_S$ given in (\ref{f5}), one can prove that
\begin{eqnarray}
\label{f6}
{\hat\partial}_i D^{\alpha\beta}_S(\tau)&=&-4\;\frac{\;P_{ij}r^j}
{r-{\bf v}\cdot{\bf r}}\;\frac{r_{\gamma} S^{\gamma(\alpha}u^{\beta)}}
{(r-{\bf k}\cdot{\bf r})^2}+
4\int^{s}_{-\infty}\frac{P_{ij}S^{j(\alpha}u^{\beta)}}{(r-{\bf k}\cdot{\bf r})^2}
\;d\zeta+8
\int^{s}_{-\infty}\frac{P_{ij}r^j\;
k_\gamma S^{\gamma(\alpha}u^{\beta)}}
{(r-{\bf k}\cdot{\bf r})^3}\;d\zeta\;.
\end{eqnarray}

The remaining integrals in (\ref{f5}) and (\ref{f6}) have the
following form
\begin{equation}
\label{f7a}
{\cal I}(s)=\int^s_{-\infty}\frac{1-{\bf k}\cdot{\bf v}(\zeta)}
{(r-{\bf k}\cdot{\bf r})^n}\;F(\zeta)\;d\zeta\;,
\quad\quad\quad\quad (n=2,3)
\end{equation}
where $F(\zeta)$ is a smooth function of the 4-velocity and/or
spin of the bodies at the retarded time. The integral (\ref{f7a})
can be calculated by making use of a new variable (cf. equation (\ref{sd5}))
\begin{equation}
\label{new1}
y={\bf k}\cdot{\bf r}-r=\zeta-t^{\ast}-{\bf k}\cdot {\bf z}(\zeta)\;,\quad\quad\quad\quad
dy=\left[1-{\bf k}\cdot {\bf v}(\zeta)\right]d\zeta\;,
\end{equation}
so that the above integral (\ref{f7a}) can be expressed as
\begin{eqnarray}
\label{f7b}
{\cal I}(s)&=&(-1)^n\int^{y(s)}_{-\infty}\frac{F(y)\;dy}{y^n}=
{(-1)^{n+1}\over n-1}\int^{y(s)}_{-\infty}F(y){d\over dy}\left({1\over y^{n-1}}\right)dy\;.
\end{eqnarray}
Integration by parts results in
\begin{eqnarray}
\label{f7c}
{\cal I}(s)&=&{1\over n-1}\frac{F(y(s))}{(r-{\bf k}\cdot{\bf r})^{n-1}}-
{1\over n-1}\int^s_{-\infty}\frac{\dot{F}(\zeta)\;d\zeta}
{(r-{\bf k}\cdot{\bf r})^{n-1}}\;.
\end{eqnarray}
The last integral in (\ref{f7c}) can be neglected under ordinary
circumstances for it includes the acceleration and/or a time
derivative of the spin of the body. Omitting such terms in accordance with our general
approximation scheme, we finally arrive at
\begin{eqnarray}
\label{f5a}
B^{\alpha\beta}_S(\tau)&=&
\frac{4r_\gamma S^{\gamma(\alpha}u^{\beta)}}{(r-{\bf v}\cdot{\bf r})
(r-{\bf k}\cdot{\bf r})}-
{4\over 1-{\bf k}\cdot{\bf v}}
\frac{k_\gamma S^{\gamma(\alpha}u^{\beta)}}{(r-{\bf k}\cdot{\bf r})}\;,
\\\nonumber\\
\label{f6a}
{\hat\partial}_i D^{\alpha\beta}_S(\tau)&=&-\;\frac{4\;P_{ij}r^j}
{r-{\bf v}\cdot{\bf r}}\;\frac{r_{\gamma} S^{\gamma(\alpha}u^{\beta)}}
{(r-{\bf k}\cdot{\bf r})^2}+
{4\;P_{ij}r^j\over 1-{\bf k}\cdot{\bf v}}
\frac{k_\gamma S^{\gamma(\alpha}u^{\beta)}}
{(r-{\bf k}\cdot{\bf r})^2}+
{4\;P_{ij}v^j\over (1-{\bf k}\cdot{\bf v})^2}
\frac{k_\gamma S^{\gamma(\alpha}u^{\beta)}}
{(r-{\bf k}\cdot{\bf r})}+
{P_{ij}S^{j(\alpha}u^{\beta)}\over (1-{\bf k}\cdot{\bf v})(r-{\bf k}\cdot{\bf r})}
\;.
\end{eqnarray}

\section{Auxiliary Algebraic and Differential Relationships}

In this appendix we give several algebraic relationships that can be used for the
calculation of observable effects. Making use of (\ref{tgh}) and definition
(\ref{spvec}) we have
\begin{eqnarray}
\label{f7}
S^{i0}&=&\gamma ({\bf v}\times{\bm{\cal J}})^i\;,
\end{eqnarray}
\begin{eqnarray}
\label{f8}
S^{ij}&=&\gamma\varepsilon_{ijk}{\cal J}^k+\frac{1-\gamma}{v^2}({\bf v}
\cdot{\bm{\cal J}})\varepsilon_{ijk}v^k\;,
\end{eqnarray}
\begin{eqnarray}
\label{f9}
k_\alpha r_\beta S^{\alpha\beta}&=&\gamma\left[{\bm{\cal J}}\cdot({\bf
k}\times{\bf r})+{\bm{\cal J}}\cdot(({\bf r}-{\bf k}r)\times{\bf v})
\right]+\frac{1-\gamma}{v^2}({\bf
v}\cdot{\bm{\cal J}})({\bf k}\times{\bf r})\cdot{\bf v}\;,
\end{eqnarray}
where $\gamma=(1-v^2)^{-1/2}$ and $\varepsilon_{ijk}\equiv\epsilon_{0ijk}$ \cite{7}.

In addition, one has the following formulas of differentiation with respect to the impact parameter $\xi^i$ (a dot
over any quantity denotes differentiation with respect to time $t$)
\begin{eqnarray}
\label{poo}
\hat{\partial}_i r&=&\frac{P_{ij}r^j}
{r-{\bf v}\cdot{\bf r}}\;,\qquad\quad \hat{\partial}_i\equiv P_{ij}\frac{\partial}{\partial\xi^j}\;,
\end{eqnarray}
\begin{eqnarray}
\label{poo1}
\hat{\partial}_i r^j&=&P_{ij}+\frac{v^j P_{ik}r^k}
{r-{\bf v}\cdot{\bf r}}\;,
\end{eqnarray}
\begin{eqnarray}
\label{poo2}
\hat{\partial}_i\left(r_\beta S^{\alpha\beta}\right)
&=&P_{ij}S^{\alpha j}-\frac{r_\beta \dot{S}^{\alpha\beta}P_{ij}r^j}
{r-{\bf v}\cdot{\bf r}}\;,
\end{eqnarray}
\begin{eqnarray}
\label{po1}
\hat{\partial}_i\left(\frac{1}{r-{\bf k}\cdot{\bf
r}}\right)
&=&-\frac{(1-{\bf k}\cdot{\bf v})P_{ij}r^j}{(r-{\bf v}\cdot{\bf r})
(r-{\bf k}\cdot{\bf r})^2}\;,
\end{eqnarray}
\begin{eqnarray}
\label{po}
\hat{\partial}_i\left(\frac{1}{r-{\bf v}\cdot{\bf
r}}\right)&=&-\frac{(1-v^2+{\bf a}\cdot{\bf r})P_{ij}r^j}{(r-{\bf v}\cdot{\bf r})^3}+
\frac{P_{ij}v^j}{(r-{\bf v}\cdot{\bf r})^2}\;,
\end{eqnarray}
where ${\bf a}=\dot{\bf v}$ is the light-ray deflecting body's
acceleration and it has been assumed that ${\bf r}={\bf x}-{\bf
z}(s)={\bf k}\tau+{\bm{\xi}}-{\bf z}(s)$.

\newpage
\begin{figure*}
\centerline{\psfig{figure=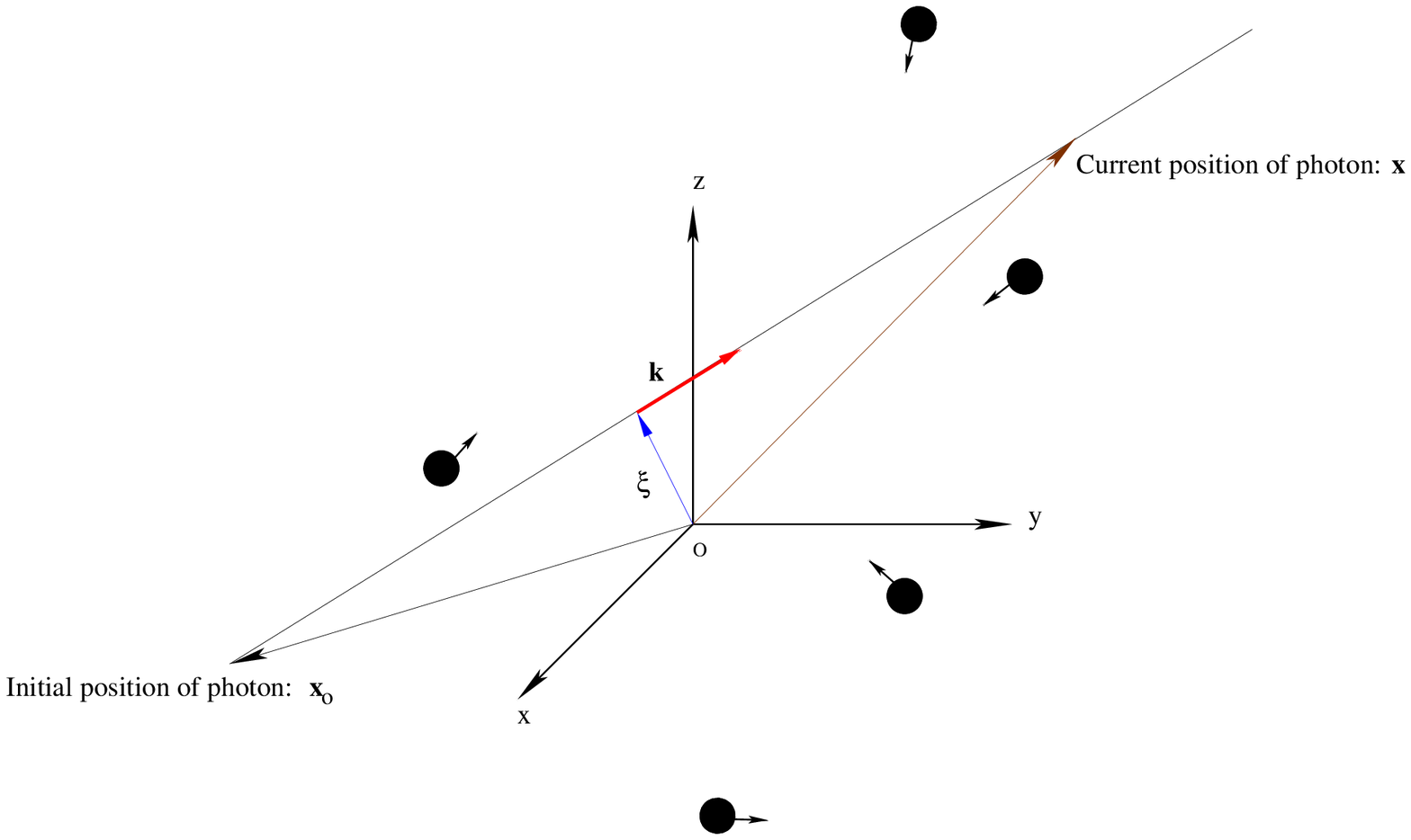,angle=0,height=21cm,width=21cm}}
\vspace{0.5cm}
\caption{Astronomical coordinate
system used for the calculations. The origin of
the coordinate system is at an arbitrary point in space. The unperturbed trajectory of a light ray is defined by the unit vector ${\bf k}$ directed from the source
of light towards the observer. The impact parameter of the light ray is defined by the vector ${\bm\xi}$ which is orthogonal to ${\bf k}$. Gravitating bodies with spin move along arbitrary world-lines.}
\label{spinfig1}
\end{figure*}


\begin{references}
\bibitem{grwave} Exhaustive consideration of the problem of interaction of quadrupolar gravitational waves
with electromagnetic signals in the approximation of geometric optics is given in \cite{KSGE} and \cite{KS}, where results obtained by previous authors are critically discussed and examined.
\bibitem{will} The speed of propagation of gravitational interaction can be different from the speed of light in alternative theories of gravitation. For more details see \cite{w}.
\bibitem{wecal} This effect was first discussed by Skrotskii \cite{Sk57}. The most exhaustive treatment of the effect in the
framework of the post-Newtonian approximation has been given recently by Klioner \cite{kl11}.
\bibitem{quad} For example, the relativistic perturbations of the ray trajectory
caused by the time-dependent quadrupole moment of a
self-gravitating body at rest have been analyzed in \cite{KSGE}.
\bibitem{2} Throughout this paper, we use units such that
Newton's gravitational constant, $G$, and the speed of light, $c$,
equal unity.
\bibitem{7} We use the convention $\epsilon_{0123}=+1$. It is worthwhile to
note that $\epsilon^{0ijk}=-\epsilon_{0ijk}$ and $\epsilon_{0ijk}=
\varepsilon_{ijk}$, where $\varepsilon_{ijk}$ is the
three-dimensional fully antisymmetric symbol such that
$\varepsilon_{ijk}= \varepsilon^{ijk}$ and $\varepsilon_{123}=+1$.
\bibitem{propa} Equation (\ref{fgh2}) evidently implies that gravity propagates with the speed of light $c=1$ in the first post-Minkowskian approximation.
\bibitem{9} The harmonic gauge is fixed in the first post-Minkowskian
approximation by the four differential conditions
\begin{eqnarray}
\nonumber
h_{\alpha}^{\;\beta},_{\beta}-\frac{1}{2} h^{\beta}_{\;\beta},_{\alpha}=0\;,
\end{eqnarray}
where a comma denotes partial differentiation.
\bibitem{genfor} See, for example, the textbooks \cite{MTW} and
\cite{penrose}, a classical review \cite{frolov}, and references therein.
Derivation and proofs of equations of the present section are also given
in Appendix A.
\bibitem{index} The tetrad indices run from 0 to 3 and indicate
time and space coordinates. The indices enclosed in the round
brackets indicate the different vectors of the tetrad and also run
from 0 to 3. It is worth noting  that time-space indices of the
tetrad have no relation to the other tetrad indices. If one
changes space-time coordinates then each tetrad vector changes in
accordance with the transformation law for vectors. On the other
hand one, can change the basis vectors at each point in space-time
by a Lorentz transformation without changing the coordinate chart
\cite{tetrad}.
\bibitem{10} The distance $X$ is usually much smaller than the distance
between the pulsar and its companion and is about 5-10 times the radius of a neutron star.
\bibitem{aij} It is worth noting that, strictly speaking, the tensor
$\hat{a}_{ij}$ can not be constant. It falls off to zero as the distance from the
source of the gravitational waves tends to infinity. However, one may assume that $\hat{a}_{ij}$ is approximately constant if
the distance, $D$, between points of
emission and observation of the electromagnetic wave is much smaller
than the distance, $r$, to the source of the gravitational wave and the
distance, $D$, is at the same time much smaller than the wavelenght $\lambda_{gr}$
of the gravitational wave.
\bibitem{note3} It is well-known that equation (\ref{m1a})
can be written in terms of usual (non-covariant) derivatives
$\partial_\alpha=\partial/\partial x^\alpha$. This is due to the complete
antisymmetry of the equation with respect to the cyclic permutation of all
indices.
\end{references}
\end{document}